\documentclass[aps, nofootinbib,superscriptaddress, preprintnumbers, longbibliography]{article}

\usepackage{tikz}
\usepackage{jcappub}
\usepackage[utf8]{inputenc}
\usepackage{amsmath,amsfonts,amssymb}
\usepackage{graphicx}
\usepackage[toc,page]{appendix}
\usepackage{cleveref}
\usepackage[loose]{units}
\usepackage{booktabs}
\usepackage{mathtools}
\newcommand{\beqn}{\begin{equation}}
\newcommand{\eeqn}{\end{equation}}

\def\la{~\mbox{\raisebox{-.6ex}{$\stackrel{<}{\sim}$}}~}

\makeatletter
\gdef\@fpheader{}
\makeatother

\begin{document}

\title{
Unveiling the nonlinear dynamics of a rolling axion during inflation
}

\author[a]{Angelo Caravano,}
\author[b,c]{Marco Peloso}

\affiliation[a]{Institut d'Astrophysique de Paris, UMR 7095 du CNRS et de Sorbonne Universit\'e,\\ 98 bis bd Arago, 75014 Paris, France}

\affiliation[b]{Dipartimento di Fisica e Astronomia “G. Galilei”, Universit\`a degli Studi di Padova, via
Marzolo 8, I-35131 Padova, Italy}
\affiliation[c]{
INFN, Sezione di Padova, via Marzolo 8, I-35131 Padova, Italy}

\emailAdd{caravano@iap.fr, marco.peloso@pd.infn.it}

\abstract{A spectator axion-gauge sector, minimally coupled to the inflaton, with the axion experiencing a momentary stage of fast roll during cosmological inflation, can generate unique signatures in primordial density fluctuations and the gravitational wave background. We present the first lattice simulation of this system using a novel hybrid numerical scheme. This approach solves the fully nonlinear dynamics of the axion-gauge sector while treating the gravitational interaction between the axion and inflaton linearly. Initially, we test the validity of the WKB approximation in the linear regime. We then investigate strong backreaction dynamics within the axion-gauge sector. Our findings reveal that backreaction significantly suppresses the growth of the gauge field and the amplitude of scalar perturbations. The simulation also allows us to analyze the non-Gaussianity of scalar fluctuations, including higher-order statistics. We show that, although non-Gaussianity is suppressed by strong backreaction, it remains higher than in the minimal model where the axion coincides with the inflaton. Our results highlight the need for simulations to make robust predictions to test against data from gravitational wave interferometers and large-scale structure surveys.
}

\maketitle

\section{Introduction}
\label{sec:introduction}

Cosmological inflation~\cite{PhysRevD.23.347,Sato:1980yn,Linde:1981mu,PhysRevLett.48.1220,STAROBINSKY198099} explains the properties of the universe at large scales and provides a mechanism for the generation of the primordial perturbations~\cite{Starobinsky:1979ty,Mukhanov:1981xt,HAWKING1982295,PhysRevLett.49.1110,STAROBINSKY1982175,Abbott:1984fp}, with excellent agreement with observations~\cite{Planck:2018jri,Planck:2019kim,BICEP:2021xfz}. A large experimental effort is devoted to the search for the (still elusive) tensor signal from inflation. Under the standard assumption that this signal is generated by the inflationary expansion, its amplitude will indicate the energy scale at which inflation took place, which is probably the most important unknown quantity in this framework. This is no longer the case if additional sources produce a tensor signal greater than the one from the expansion. Examples of this from the literature include the generation of additional gravitational waves (GW) from spectator fields~\cite{Biagetti:2013kwa,Biagetti:2014asa,Fujita:2014oba}, from an effective field theory approach of broken spatial diffeomorphisms~\cite{Cannone:2014uqa}, or from particle and string production during inflation~\cite{Cook:2011hg,Senatore:2011sp,Barnaby:2012xt,Carney:2012pk}. Such mechanisms are limited by the fact that these sources also produce scalar perturbations, typically exceeding the generation of tensor modes. This results in many of these models having a decreased overall tensor-to-scalar ratio and makes it impossible to observe the produced gravitational waves (GW) once the limits from scalar production, enforced by constraints on non-Gaussianity at Cosmic Microwave Background (CMB) scales or on Primordial Black Hole (PBH) abundance, are taken into account~\cite{Biagetti:2014asa,Barnaby:2012xt,Papageorgiou:2019ecb}.

The situation is better under specific circumstances. For instance, a decreased ratio of sourced scalar vs. sourced tensor perturbations is produced if the sourcing fields are coupled only gravitationally to the inflaton and if they are relativistic (as a nonrelativistic source has a suppressed quadrupole moment, and thus a reduced GW production~\cite{Barnaby:2012xt}). These two features are present in the model of Refs.~\cite{Barnaby:2012xt,Mukohyama:2014gba}, in which massless U(1) gauge fields are sourced by their axial coupling to a spectator axion. This mechanism was previously studied in the context in which the rolling axion field was the inflaton, for which, as already mentioned, stronger bounds from the scalar production are present~\cite{Barnaby:2010vf,Sorbo:2011rz,Barnaby:2011vw,Barnaby:2011qe,Linde:2012bt}.~\footnote{See also Ref. \cite{Fujita:2022fit} for a more recent treatment of this model using the stochastic formalism.} In these works, the axion is assumed to continuously roll all throughout inflation. Ref.~\cite{Namba:2015gja} considered instead a spectator axion with a mass of the order of the Hubble rate, that rolls in a typical cosine potential for only a few $e$-folds (typically, $\sim 2 -5$) of inflation. This results in sourced signals localized at the scales that left the horizon while the axion was rolling. This significantly decreases the limits from the CMB non-Gaussianity, so that this mechanism can produce an observable tensor signal at CMB scales for an arbitrarily low scale of inflation~\cite{Namba:2015gja}.

From the theory point of view, axions are ubiquitous in string theory, and for example the recent work~\cite{Dimastrogiovanni:2023juq} studied how this mechanism might be realized in string compactifications (where in fact a multiplicity of axions might roll at different stages during inflation, producing a multitude of peaked signals). From a phenomenological point of view, the mechanism of~\cite{Namba:2015gja} for the generation of localized signals has been employed in a number of works that directly studied it or discussed it in the context of the CMB~\cite{Ozsoy:2017blg,NASAPICO:2019thw,Shandera:2019ufi,Planck:2019kim,Campeti:2019ylm,Abazajian:2019eic,Komatsu:2022nvu,Campeti:2022acx,Fujita:2022qlk,LiteBIRD:2023zmo} (including CMB spectral distortions~\cite{Putti:2024uyr}), of PBH dark matter~\cite{Garcia-Bellido:2016dkw,Garcia-Bellido:2017aan,Ozsoy:2020kat,Talebian:2022cwk,Ozsoy:2023ryl,LISACosmologyWorkingGroup:2023njw}, and of GW generation at smaller scales~\cite{Caprini:2019pxz,Giare:2020vhn,Campeti:2020xwn,Figueroa:2022iho,LISACosmologyWorkingGroup:2022jok,Giare:2022wxq,Corba:2024tfz,Ozsoy:2024apn}, also in relation to the recent evidence for a stochastic gravitational wave background (SGWB) from pulsar timing array (PTA) measurements~\cite{NANOGrav:2023hvm,Figueroa:2023zhu,Unal:2023srk,Jiang:2023gfe,Fu:2023aab,Niu:2023bsr}.\footnote{These mechanisms can be readily extended to the SU(2) case. Refs.~\cite{Dimastrogiovanni:2016fuu} and~\cite{Iarygina:2021bxq} studied the non-abelian counterparts of~\cite{Barnaby:2012xt,Mukohyama:2014gba}, with gauge field production in a spectator sector that resembles the inflationary sector of, respectively, chromo-natural inflation~\cite{Adshead:2012kp} and gauge-flation~\cite{Maleknejad:2011jw}. The possibility of a localized axion roll in this context~\cite{Thorne:2017jft} is technically more difficult than in the U(1) case, and it can be implemented for instance with a varying coupling between the axion and the gauge field~\cite{Putti:2024uyr}.} Notably, in this context, precision astrometry will allow us to probe the chirality of the SGWB background~\cite{Liang:2023pbj}. This will allow us to determine whether the PTA signal can be explained by axions during inflation, as the SGWB generated by axionic interaction is parity-violating. Moreover, the parity-odd nature of axion-gauge interactions generates a sizable parity-odd scalar primordial trispectrum \cite{Fujita:2023inz,Niu:2022fki}, which could explain the recent claim of a parity-odd 4-point correlation function of galaxies from the BOSS survey~\cite{Hou:2022wfj,Philcox:2022hkh}.~\footnote{A recent study~\cite{Philcox:2024mmz} has shown that an alternative covariance estimation reduces this signal. Consequently, the statistical significance of this signal is still an open question, as explained for example in~\cite{Jamieson:2024mau}.} 

In light of the phenomenological interest for these models, an extremely relevant topic of discussion is the backreaction of the produced gauge fields on the background evolution of the axion, and the validity of the perturbative computation with which these scalar and tensor signals are computed~\cite{Ferreira:2015omg,Peloso:2016gqs,Dimastrogiovanni:2024lzj}. The analytical estimates of~\cite{Peloso:2016gqs} indicated that backreaction and perturbativity are under control for $ \xi \lesssim 5 $, where $ \xi $ is the typical parameter that controls the gauge field amplification, defined in eq.~(\ref{eq:xi}) below. This is close to the value required for several phenomenological applications of this mechanism. Therefore, while the analytic methods of~\cite{Namba:2015gja} can provide a reasonable order of magnitude estimate of these signals, numerical computations appear to be necessary to provide more precise answers. The present work is the first numerical study of the rolling axion model of Ref.~\cite{Namba:2015gja} using nonlinear lattice simulations.

\begin{figure}
	\centering
	\includegraphics[scale=0.45]{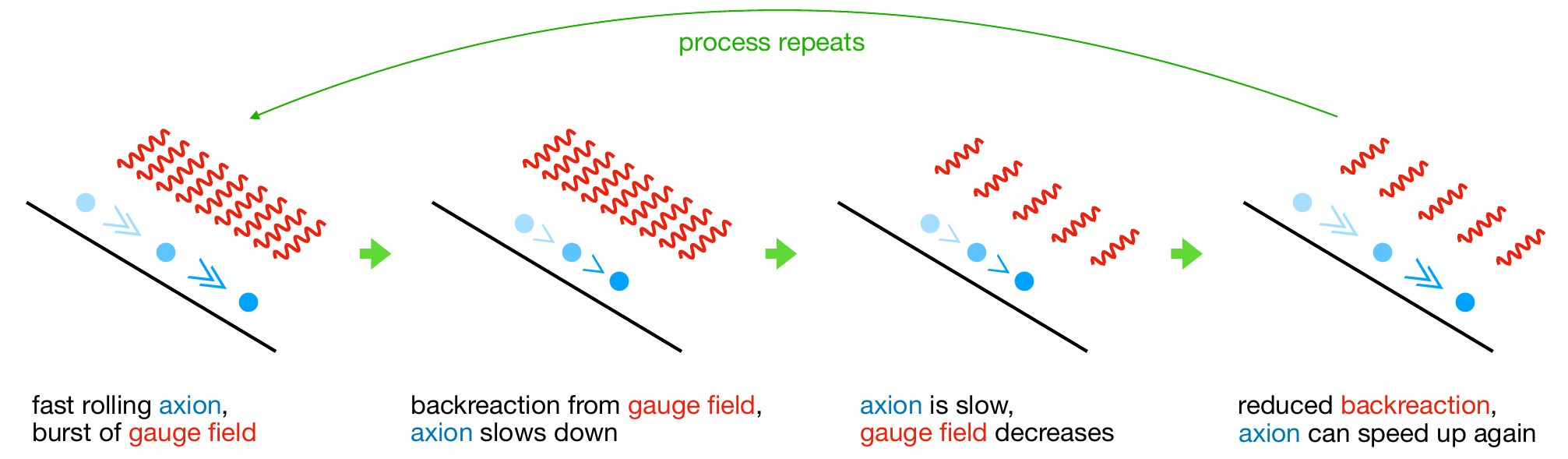}
	\caption{Intuitive illustration of the cyclic backreaction process in axion inflation, explained in \cref{sec:introduction}.}
	\label{fig:illustration}
\end{figure}

Understanding backreaction is a central focus in the recent literature. In Ref.~\cite{Anber:2009ua}, Anber and Sorbo (AS) have provided a very simple and minimal implementation of warm inflation~\cite{Berera:1995ie}. In their solution, the axion inflaton has a steady-state speed, for which the friction due to backreacting gauge fields balances the slope of the axion potential. This friction can be greater than the usual Hubble friction term, allowing for an inflationary solution in potentials that would be too steep to sustain inflation in absence of gauge field production. A number of studies have shown that the AS steady-state solution is unstable. Most of these works solved, either analytically~\cite{Peloso:2022ovc} or numerically~\cite{Cheng:2015oqa,Notari:2016npn,DallAgata:2019yrr,Domcke:2020zez,Gorbar:2021rlt,Durrer:2023rhc,vonEckardstein:2023gwk,Iarygina:2023mtj} the evolution equations of this model for a homogeneous axion coupled to gauge fields (namely, under the working assumption of neglecting the inhmogeneties of the axion).~\footnote{See also Ref.~\cite{Galanti:2024jhw} for a recent work where backreaction is computed with a perturbative scheme.} It was understood in~\cite{Domcke:2020zez} (and then confirmed by the analytic kernel present in the computation of Ref.~\cite{Peloso:2022ovc}) that the instability is due the fact that the amplitude to which each gauge field mode contributes to backreaction at any time $t$ is sensitive to the values of the axion velocity at earlier times $t'<t$, while that mode was being amplified. This introduces a memory effect, to which backreaction tries to `adapt' (rather than the instantaneous balance between the potential slope and the dissipation).~\footnote{Instantaneous backreaction can be achieved in the construction of Ref.~\cite{Creminelli:2023aly}.} This results in an oscillatory behaviour of the axion velocity about the AS solution: sudden episodes of faster axion speed result in bursts of gauge field amplification, and consequent backreaction, that slow down the velocity of the axion, so that the gauge field production strongly decreases, reducing backreaction, giving rise to the next episode of fast roll where the gauge field grows again. This oscillatory behaviour, illustrated in \cref{fig:illustration}, has a period of $\sim 4-5$ $e$-folds~\cite{Domcke:2020zez,Peloso:2022ovc}, and it could result in correlated features in the primordial GW spectrum at different scales~\cite{Garcia-Bellido:2023ser}.

Recently, lattice simulations have started to improve our knowledge of the signatures of gauge fields in axion inflation, confirming the validity of earlier analytic results in the small backreaction regime, and providing new insight into the complementary regime of strong backreaction. Lattice simulations are a well-known tool in early-Universe cosmology, and typically employed to study nonlinear cosmological phenomena such as cosmological phase transitions and reheating (see, e.g.,~\cite{Khlebnikov_1996,Prokopec_1997,latticeeasy,Garcia-Bellido:2007nns,Frolov_2008,hlattice,Sainio_2012,Child_2013,Easther_2010,Lozanov_2020,figueroa2021cosmolattice,Adshead:2016iae,Adshead:2017xll}). More recently, these numerical techniques have been extended to the inflationary epoch and are emerging as a crucial tool in this context~\cite{Caravano:2021pgc,Caravano:2021bfn,Caravano:2022epk,Caravano:2022yyv,Krajewski:2022uuh,Figueroa:2023oxc,Caravano:2024tlp}.

Lattice simulations can take backreaction fully into account, relaxing the assumptions of previous analytic and numerical studies (such as neglecting fluctuations in the axion field). The first lattice simulation of axion inflation was presented in Ref.~\cite{Caravano:2022epk}. This work confirmed the instability of the AS solution and the oscillatory behavior, showing only the presence of the first oscillatory peak in the axion velocity due to limited UV resolution. Subsequently, Ref.~\cite{Figueroa:2023oxc} showed that these oscillations are partially smoothed out during the final few $e$-folds of inflation. In light of these results, the validity of a repeated cyclic backreaction, depicted in \cref{fig:illustration}, remains an open question. Indeed, these studies represent initial steps towards a comprehensive understanding of backreaction in this system. A central challenge is the UV sensitivity of the minimal axion inflation setup, where the axion coincides with the inflaton. In this scenario, the gauge field is excited across all scales until the end of inflation, necessitating costly lattice simulations with a large dynamical range. Furthermore, in the regime near the end of inflation studied in Ref.~\cite{Figueroa:2023oxc}, metric fluctuations, usually neglected in lattice simulations, might become significant, requiring numerical relativity techniques to solve the full General Relativity equations along with nonlinear field equations, as recently explored in \cite{Adshead:2023mvt}.\footnote{An alternative approach would be treating gravity linearly, and check afterwards that metric perturbations remain small. This is precisely the approach that we use in this work to study the gravitational interaction between the spectator axion and the inflaton.}

Lattice studies have so far focused on the minimal setup where the axion coincides with the inflaton. As explained above, this scenario requires a large dynamical range due to the gauge field being enhanced at all scales, necessitating costly lattice simulations. However, if the axion acts as a spectator particle, rolling for only a few $e$-folds, the required dynamical range to study the nonlinear dynamics of this system is significantly reduced. The fast-rolling model thus presents an ideal environment to investigate backreaction in axion inflation and its phenomenological implications. In this work, we present the first lattice study of a spectator axion-gauge system, coupled to the inflaton through gravity. To achieve this, we introduce a novel numerical scheme that solves the spectator axion-gauge sector in a fully nonlinear fashion (i.e. as in Refs.~\cite{Caravano:2022epk,Figueroa:2023oxc}), while simultaneously addressing the gravitational interaction between the spectator sector and the axion in a linear way.~\footnote{A similar approach was followed in Refs.~\cite{Adshead:2018doq,Adshead:2019igv,Adshead:2019lbr} that computed GW production during gauge preheating by adding tensor modes linearly to a lattice computation.}

This paper is organized as follows. In \cref{sec:linear}, we provide a brief review of the perturbative (analytical) treatment of this model. In \cref{sec:simulation}, we introduce our hybrid nonlinear scheme used to solve this system on the lattice. In \cref{sec:weak} and \cref{sec:results_strong}, we present results from lattice simulations in regimes of small and strong backreaction, respectively. Finally, in \cref{sec:conclusion}, we discuss our results and outline future directions. The work is concluded by \cref{app:sigma}, in which we present the analytic computation of the axion perturbations (providing results not explictly given in~\cite{Namba:2015gja}) and by \cref{app:resolution}, in which we show some tests of our lattice computations.

\section{Review of the linear theory}
\label{sec:linear}
\subsection{The model}

The model~\cite{Namba:2015gja} is characterized by an inflaton field $\phi$ and an axion $\sigma$ that are coupled to each other only gravitationally. The axion has a typical cosine potential, and a standard pseudo-scalar interaction with a U(1) field, so that the lagrangian is 
\begin{equation}
{\cal L} = - \frac{1}{2} \left( \partial \phi \right)^2 - V \left( \phi \right)  - \frac{1}{2} \left( \partial \sigma \right)^2 - \frac{\Lambda^4}{2} \left[ \cos \left( \frac{\sigma}{f} \right) + 1 \right]  - \frac{1}{4} F^2 - \frac{\alpha}{4 f} \sigma F {\tilde F},
\label{lag}
\end{equation}
where $F_{\mu \nu} \equiv \partial_\mu A_\nu - \partial_\nu A_\mu$ is the field strength of the U(1) field, and ${\tilde F}^{\mu \nu} \equiv \frac{\epsilon^{\mu \nu \rho \sigma}}{2\sqrt{-g}} F_{\rho \sigma}$ is its dual.~\footnote{The quantity $g$ is the determinant of the metric, while $\epsilon^{\mu \nu \rho \sigma}$ is totally anti-symmetric, with $\epsilon^{0123} = 1$. We assume a FLRW geometry, with line element $ds^2 = a^2 \left( \tau \right) \left[ - d \tau^2 + d \vec{x}^2 \right]$, where $a$ is the scale factor and $\tau$ the conformal time.} The parameter $f$ is denoted as the axion decay constant and it has mass dimension one. The coefficient $\alpha$ controls the coupling to the gauge field and it is dimensionless. The discussion in this section is independent of the specific inflaton potential, which is only assumed to be flat enough to lead to slow roll. The specific potential employed in our simulations is described in Subsection~\ref{subsubsec:bck}.

The axion mass is tuned to be of the same order as the Hubble rate $H$ during inflation, $\delta \equiv \Lambda^4 / \left( 6 H^2 f^2 \right) \la 1$. Then, for $\delta \ll 3$, and to leading order in slow roll~\cite{Namba:2015gja}
\begin{equation}
\dot{\sigma} = \frac{f H \delta}{\cosh \left[ \delta H \left( t - t_* \right) \right]},
\label{sigma-dot}
\end{equation}
where dot denotes derivative with respect to (wrt) physical time $t$, while $t_*$ is the time at which $\dot{\sigma}$ reaches its maximum value, that, to leading order in slow-roll, occurs when $\sigma_* = \frac{\pi}{2} f$, namely when the axion is in the steepest part of the potential. Throughout this work, we use the subscript ``$*$" to denote quantities evaluated at the time $t_*$. The expression~(\ref{sigma-dot}) shows that the axion roll is significant only for a number of e-folds $\Delta N \simeq \frac{1}{\delta}$ around this time. The motion of the axion can results in a large gauge field amplification, whenever $\xi > 1$, where~\cite{Namba:2015gja}  
\begin{equation}
\label{eq:xi}
\xi \equiv \frac{\alpha \, \dot \sigma}{2 H f} = \frac{\xi_*}{\cosh \left[ {\delta H \left( t-t_* \right)} \right]}=\frac{2 \, \xi_*}{ \left( \frac{a}{a_*} \right)^\delta + \left( \frac{a_*}{a} \right)^\delta }, 
\end{equation}
where $\xi_* = \frac{\alpha \, \delta}{2}$ is the maximum value assumed by $\xi$ at the time $t = t_*$. We stress that these analytic results are given to leading order in slow-roll and neglecting backreaction from the gauge fields amplified by the motion of the axion. In particular, they assume a constant Hubble rate, and therefore the scale factor evolution $a = - \frac{1}{H\tau}$ during inflation (de Sitter geometry). In Section~\ref{sec:weak}, these analytic results are compared with the full numerical solutions from the lattice. 

\subsection{Gauge field production}

The equation of motion for the gauage field mode function following from~\cref{lag} is~\cite{Namba:2015gja}  
\begin{equation}
\label{eq:Alin}
    A^{\prime\prime}_{\pm}+\left(k^2\pm \frac{4 k \xi_*}{\tau\left[ \left( \frac{\tau}{\tau_*} \right)^\delta + \left( \frac{\tau_*}{\tau} \right)^\delta \right]    }\right)A_{\pm}=0,
\end{equation}
where the axion is assumed to be homogeneous and \cref{eq:xi} has been used for the axion time derivative. Prime denotes differentiation wrt conformal time $\tau$. Ref.~\cite{Namba:2015gja} obtained an approximate anlaytic solution for the gauge mode function using the WKB method:

\begin{equation}
\label{eq:WKB}
A_{+}\simeq\left(\frac{-\tau}{8k\xi}\right)^{1/4} \tilde A(\tau,k),\quad A^\prime_{+}\simeq\left(\frac{k\xi}{-2\tau}\right)^{1/4} \tilde A(\tau,k),
\end{equation}
where:
\begin{equation}
\tilde A(\tau,k) \equiv N[\xi_*,k,\delta] \exp \left[ - \frac{4\sqrt{\xi_*}}{1+\delta} \left(\frac{\tau}{\tau_*}\right)^{\delta/2} \sqrt{-k\tau}\right],
\label{A-wkb}
\end{equation}
and $N[\xi_*,k,\delta]$ is a time-independent constant obtained, for every $k$, by matching the WKB solution for $A_+$ with the numerical solution of \cref{eq:Alin} at late times $-k\tau\ll 1$. 

The amplified gauge field sources perturbations $\delta \sigma$ of the axion field by their direct coupling in \cref{lag}. These modes are then gravitationally coupled to (scalar and tensor) pertrubations of the metric, as well as to the perturbations of the inflaton field.

\subsection{Inflationary scalar perturbations}
\label{sec:perturbations}

We work in the spatially flat gauge, for which (disregarding vector and tensor perturbations) the line element reads
\begin{equation}
d s^2 = a^2 \left( \tau \right) \left[ - \left( 1 + 2 \Phi \right) d \tau^2 + 2 \partial_i B d x^i d \tau + \delta_{ij} d x^i dx^j \right] \;,
\end{equation}
where the spatial components of the metric are unperturbed, while the scalar modes $\Phi$ and $B$ are non-dynamical and can be integrated out. Namely, these modes enter in the full action $S = \int d^4 x \sqrt{-g} \left[ \frac{M_p^2}{2} R + {\cal L} \right]$ (where $R$ is the scalar curvature and ${\cal L}$ is given in eq.~(\ref{lag})) without time derivatives. Their equations of motion in momentum space are therefore constraint equations (without derivatives acting on these modes) that admit algebraic solutions for $\Phi$ and $B$ in terms of the dynamical modes $\delta \phi$ and $\delta \sigma$. We insert these solutions back into the action, which therefore becomes an action for solely the two dynamical modes. To obtain the linear equations for these modes, plus their coupling to the vector fields, we retain in this action terms that are quadratic in the dynamical modes, plus terms of ${\rm O } \left( \delta \phi \, A \right)$ and of  ${\rm O } \left( \delta \sigma \, A \right)$. Finally we rescale the dynamical modes as $Q_\phi \equiv a \, \delta \phi$ and $Q_\sigma \equiv a \, \delta \sigma$, as these combinations are the canonical modes of this system (their kinetic term obtained from the procedure just outlined is $S_{\rm kin} = \frac{1}{2} \int d^4 x \left[ Q_\phi^{'2} + Q_\sigma^{'2} \right])$. Extremizing this action (see Section 3.1 of~\cite{Namba:2015gja} for furhter details), gives 

\begin{align}
\label{eq:lin}
&\left(\frac{\partial^2}{\partial\tau^2}+k^2+\tilde M^2_{\phi\phi}\right)Q_{\phi}+\tilde M^2_{\phi\sigma}Q_{\sigma}=0, \nonumber\\
&\left(\frac{\partial^2}{\partial\tau^2}+k^2+\tilde M^2_{\sigma\sigma}\right)Q_{\sigma}+\tilde M^2_{\sigma\phi}Q_{\phi}=\frac{\alpha a^3}{f}\int \frac{d^3x}{(2\pi)^{3/2}}e^{-i\vec{x}\cdot\vec{k}}\vec{E}\cdot\vec{B},
\end{align}
where the ``mass term'' in these relations is given by
\begin{equation}
\tilde M^2_{ij} \equiv -\frac{a^{\prime\prime}}{a}\delta_{ij}+a^2V_{,ij}+\left(3-\frac{a^2\bar\phi^\prime_k\bar\phi^\prime_k }{2a^{\prime 2}}\right)\bar\phi^\prime_i\bar\phi^\prime_j+\frac{a^3}{a^\prime}(\bar\phi^\prime_i V_{,j}+ \bar\phi^\prime_j V_{,i}),
\end{equation}
where the indices $i$ and $j$ run from $1$ ot $2$, and the value $1$ refers to $\phi$, while $2$ to $\sigma$ (with $\phi_1 \equiv \phi$, $\phi_2 \equiv \sigma$). We also note that in this relation $V$ refers to the sum of the inflaton and the axion potentials, and a comma on the potential refers to differentiation wrt the corresponding field.

The last term in \cref{eq:lin} represents the direct coupling between gauge and axion modes. The full structure of the ``mass term'' arises from the lagrangian (\ref{lag}) and from the procedure of integrating out the non-dynamical modes of the metric, which provides the gravitational interaction between the two modes (encoded in the off-diagonal $\tilde M^2_{ij}$ coefficients), as well as in their gravitational self-interactions (that contribute to the diagonal $\tilde M^2_{ij}$ coefficients).~\footnote{As detailed in \cref{sec:simulation}, only the first equation in \eqref{eq:lin} is added to the system of equations solved in the lattice simulations, since the set of equations \eqref{eq:nonlin} already incorporates the coupling between gauge fields and the axion. This means that we are neglecting the backreaction of inflaton and metric perturbations on the axion-gauge field dynamics. Additionally, in the first equation of \eqref{eq:lin}, we disregard the gravitational coupling between gauge modes and inflaton perturbations.  This coupling, whose effect was studied in Ref.~\cite{Barnaby:2012xt}, is also of gravitational origin (so, at the technical level, it is equally $M_p$ suppressed as the one we consider here), but it occurs only at horizon crossing, while the $\delta \phi - \delta \sigma$ that we consider here continues while both fields are rolling. Comparing with the results of~Ref.~\cite{Barnaby:2012xt}, Refs.~\cite{Ferreira:2014zia} (see the final part of their Section 6.1) and~\cite{Namba:2015gja} (see for their footnote 10) showed that the term considered here provides the dominant contribution.} To obtain analytical predictions, one must substitute the WKB ansatz for the gauge field, derived in the previous section, into this set of equations. In \cref{app:sigma}, we solve the second of eqs. \eqref{eq:lin} and derive analytical predictions for the power spectrum of $\sigma$. The first of eqs. \eqref{eq:lin} is analytically solved in Ref.~\cite{Namba:2015gja}, which presents results for the power spectrum of $\phi$. In \cref{sec:weak,sec:results_strong}, we will compare these analytical calculations with the results from lattice simulations.

\subsection{Backreaction and perturbativity}
The perturbative treatment outlined in this section is valid under certain assumptions, which we will relax thanks to the lattice simulation. In particular, if gauge field production is too abundant, it can affect the background dynamics of the axion field, invalidating the standard perturbative approach. Backreaction and perturbativity bounds for this model are detailed in Refs.~\cite{Campeti:2022acx,Peloso:2016gqs}, which provide a comprehensive analysis of these constraints. A recent motivation for this study is the fact that the parameter range of interest for this model to explain the recently observed gravitational wave background using PTA experiments lies in the strong backreaction regime of the theory~\cite{Unal:2023srk}. In \cref{sec:results_strong}, we will run simulations in this regime and quantify the deviations from the predictions of standard perturbation theory.

\section{Lattice simulation}
\label{sec:simulation}
To simulate the dynamics of the spectator axion-gauge system and its interaction with the inflaton, we introduce a novel hybrid numerical approach. This method addresses the fully nonlinear dynamics of the spectator axion-gauge system while treating the inflaton sector perturbatively. In this section, we first present the details of this hybrid scheme and discuss its validity and applicability. Then, we describe the procedure to generate the initial conditions for the simulation.

\subsection{Equations of motion}
\label{sec:eoms}

The lattice approach consists of discretizing continuous space onto a grid of $N_{\rm pts}^3$ points separated by the distance $dx=L/N_{\rm pts}$, where $L$ is the comoving length of the box. We then numerically solve the equations of motion in real space for each grid point, using periodic boundary conditions.
The nonlinear equations for the axion-gauge spectator sector are the following \cite{Caravano:2022epk}:
\begin{align}
		\label{eq:nonlin}
		\begin{split}
			&		\sigma^{\prime\prime}+2H{\sigma^\prime}-\nabla^2\sigma+a^2\frac{\partial V}{\partial \sigma}=-a^2\frac{\alpha}{4f}F_{\mu\nu}\tilde F^{\mu\nu}, \\
			&{A}^{\prime\prime}_0-	\nabla^2A_0=\frac{\alpha}{f}\epsilon_{ijk}\partial_k\sigma \partial_iA_j,\\
			&{A}^{\prime\prime}_i-\nabla^2A_i=\frac{\alpha}{f}\epsilon_{i jk}\sigma^\prime \partial_jA_k-\frac{\alpha}{f}\epsilon_{i jk}\partial_j\sigma (A^{\prime}_k-\partial_kA_0),
		\end{split}
\end{align}
where $i,j,k\in\{1,2,3\}$ are spatial indices, and $\nabla^2=\partial_j\partial_j$. We choose to work in the Lorenz gauge $\partial^\mu A_\mu=0$. Due to their nonlinear nature, these equations are able to capture the backreaction of the gauge field on the background dynamics of the axion. Note that, as already mentioned in \cref{sec:perturbations}, we are neglecting the backreaction of inflaton perturbations on the axion-gauge field dynamics. This assumption is justified as long as inflaton perturbations remain small, which is true for all the cases considered in this work.

The gravitational interaction between the inflaton and the axion is determined by the linear \cref{eq:lin}, that we rewrite in real space as:
\begin{equation}
\label{eq:pinf}
    \left(\frac{\partial^2}{\partial\tau^2}+\nabla^2+\tilde M^2_{\phi\phi}\right)Q_{\phi}=-\tilde M^2_{\phi\sigma}a\left(\sigma-\langle\sigma\rangle\right).
\end{equation}
The evolution of the background quantities in \cref{eq:nonlin,eq:pinf}  is determined by solving the second Friedmann equation and the Klein-Gordon equation for the background inflaton, which are, respectively:
\begin{align}
\label{eq:back}
\begin{split}
    &a^{\prime\prime}=\frac{1}{6}\left(\langle\rho\rangle-3 \langle p\rangle\right)a^3,\\
    & \bar\phi^{\prime\prime}+2\frac{a^{\prime}}{a}\bar\phi^{\prime}=-a^2\frac{\partial V(\bar\phi)}{\partial{{\bar\phi}}},
    \end{split}
\end{align}
where $\langle\rho\rangle$ and $\langle p \rangle$ are the background energy-density and pressure, which are obtained as lattice averages of all the energy and pressure contributions. Note that the gauge field does not enter the second Friedman equation explicitly (except when setting the initial conditions), as $\rho_{GF}=3 p_{\rm GF}$.

The set of \cref{eq:nonlin,eq:pinf}, evaluated at each lattice point, plus the two eqs.~(\ref{eq:back}), forms a system of $ 6N_{\rm pts}^3 + 2 $ second-order differential equations. We solve these equations using a 4th-order Runge-Kutta time integrator and the discretization technique developed in Refs.~\cite{Caravano:2021pgc,Caravano:2021bfn,Caravano:2022epk,Caravano:2022yyv}, to which we refer for detailed information on the discretization of eqs.~(\ref{eq:nonlin}). Notably, in this approach, we evolve all four components of the gauge field $ A_{\mu} $ as independent variables. Consequently, the Lorenz constraint $ \partial^\mu A_{\mu} = 0 $ is not automatically preserved on the lattice (barring infinite accuracy) and must be verified separately. This verification is addressed in \cref{app:resolution}, where we show that the Lorenz constraint is well satisfied throughout the evolution in all the cases considered in this work.

The hybrid nonlinear description presented in this section is valid if the main inflationary sector remains perturbative. Also the right-hand side of \cref{eq:pinf} needs to remain perturbative, requiring axion perturbations to remain smaller than their background $\sigma-\langle\sigma\rangle \ll \langle\sigma\rangle $.  While this is guaranteed in the linear regime of the theory, we need to check that it remains true even in the case of strong backreaction from the gauge field. We will show this in \cref{sec:results_strong}.

\subsection{Initial conditions and resolution}
\label{subsec:inital}

The initial conditions on the lattice are set perturbatively. We initialize all fields as their background values plus some space-dependent fluctuations, that we discuss separately in the following.

\subsubsection{Background} 
\label{subsubsec:bck}

In order to set the initial values for the inflaton background, we need to choose a particular slow-roll potential $V(\phi)$. We assume the following $\alpha$-attractor potential \cite{Kallosh:2013yoa,Ferrara:2013rsa}:
\begin{equation}
\label{eq:potential}
V(\phi)=\bar V\left(1-e^{-\sqrt{\frac{2}{3\alpha}}\phi}\right)^2,
\end{equation}
where we set $\bar V=1.348\times10^{-11}$ and $\alpha=0.1$. The initial values of the background inflaton and its velocity are set to $\bar\phi_{0}= 2.53$ and $\dot{\bar\phi}_{0}= -0.0043 \sqrt{\bar V}$, 
so that there are $N\simeq 52$ $e$-folds left before the end of inflation.~\footnote{We work in reduced Planck mass units $M_{\rm Pl}=1$.} This choice translates to\begin{equation}
    n_s = 0.961, \quad A_s = 2\times 10^{-9},\quad r= 4.5\times 10^{-4} \,, 
\end{equation}
for, respectively, the spectral tilt and the amplitude of the scalar primordial perturbations and for the tensor-to-scalar ratio~\cite{Planck:2018jri}. We emphasize that, although we must select a specific potential for the lattice simulation, the results presented in this work do not crucially depend on the particular form of the potential, as long as it provides the conditions to sustain slow-roll inflation.

The background value of the axion field is set to $(\sigma_0,\dot\sigma_0)=(0.89\,f,0.20\,f\sqrt{\bar V})$. \footnote{The initial values for the background inflaton, the background axion, and their time derivatives are obtained from a numerical integration of about $2$ e-folds of the exact equations for the homogeneous inflaton and axion, and for the scale factor (without including the gauge fields), that starts from slow-roll initial conditions.} The initial condition for the axion is tuned so that it reaches its maximum velocity at $N\simeq 1.84$ $e$-folds after the start of the simulation. The gauge field background is set to zero at the beginning of the simulation. The initial value of the scale factor $a_0$ is chosen so that $a=1$ when the axion reaches its maximum velocity in the simulation.

\subsubsection{Perturbations}
\label{subsubsec:perts}

Perturbations of the axion, inflaton and gauge field are initiated in the Bunch-Davies vacuum at the beginning of the simulation:
\begin{equation}
\label{eq:BD}
\delta\phi(\vec{k},\tau)=\frac{1}{\sqrt{2k}a}e^{- i {k}\tau},\quad \delta\sigma(\vec{k},\tau)=\frac{1}{\sqrt{2k}a}e^{- i {k}\tau},\quad A_{\pm}(\vec{k},\tau)=\frac{1}{\sqrt{2k}}e^{- i {k}\tau},
\end{equation}
where $A_{\pm}$ are the gauge field polarizations. The initial fluctuations are the result of a statistical random process, which sets the power spectrum equal to the Bunch-Davies one. More details regarding the initialization procedure, and how it accounts for discretrization and finite-size effects, can be found in Refs.~\cite{Caravano:2022yyv,Caravano:2021bfn}.

The lattice size is set to $N_{\rm pts}^3=256^3$ number of points and comoving length of $L=25/(a_*H_*)$. This translates to the following IR and UV comoving cutoffs for the simulation \cite{Caravano:2021pgc}:
\begin{equation}
k_{IR}=\frac{2\pi}{L}\simeq0.25\, k_*,\quad\quad k_{UV}\simeq 35\, k_*,
\end{equation}
where $k_*=a_* H_*$ is the mode exiting the horizon at $t_*$, acting as a reference scale (as just mentioned, $a_* = 1$). We recall that $t_*$ indicates the time at which the axion reaches its maximum velocity in the full lattice simulation. We run the simulation until $N=\log(a/a_*)\simeq 4$, which makes all the modes in our lattice super-horizon at the final time. In the main text, we show results from simulations with this resolution. In \cref{app:resolution} we show that our results do not depend on the spatial resolution by running simulations with different IR and UV cutoffs.

\section{Weak backreaction}
\label{sec:weak}

We now present simulation results for a scenario where the backreaction of the produced gauge quanta on the background dynamics is small, and where linear theory predictions are expected to be reliable. In this regime, our simulations can quantitatively test the validity of the WKB approximation~\eqref{eq:WKB}. Unlike the analytical predictions, our lattice calculation does not assume a constant $H$ and it does not employ the slow-roll approximation of \cref{sigma-dot} for the axion dynamics. This allows for a more accurate background solution, and ultimately for more detailed characterization of the scale-dependence of the gauge fields amplified by the axion, and for the scalar fluctuations sourced by the gauge fields. Additionally, we use the simulation to investigate the role of high-order non-Gaussianity beyond the bispectrum.

For this simulation, we set $\alpha = 20.14$, $f=0.1$ and $\delta=0.5$. This translates into $\Lambda^4=6\,\delta H^2 f^2$ and a maxmum value of $\xi=5$ in the background numerical simulation that neglects the backreaction from gauge fields.\footnote{To determine the value of $\alpha$ corresponding to $\xi_*=5$, we do not use the slow-roll approximation, which would give  $\alpha =2\,\xi_*/\delta=20$. Instead, we set a more precise value by numerically solving the background dynamics of this model starting from the slow-roll solution 2 $e$-folds before the simulation starting time, neglecting backreaction. This is analogous to what we do to set the initial background values, as explained in \cref{subsubsec:bck}.} With this choice of parameters, the backreaction of the gauge field on the background axion dynamics is expected to be moderate~\cite{Peloso:2016gqs,Campeti:2022acx}.\footnote{Ref.~\cite{Campeti:2022acx} reports perturbativity bounds for $\delta=0.6$, that we use as conservative bounds for $\delta=0.5$.}

\subsection{Background dynamics}
\begin{figure}
	\centering
	
	\begin{tikzpicture}
	\node (img) at (1.3,0) {\includegraphics[width=7.cm]{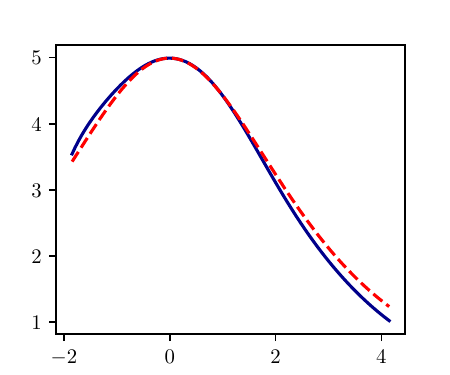}};
 \node (img2) at (-6.,0) {\includegraphics[width=7.cm]{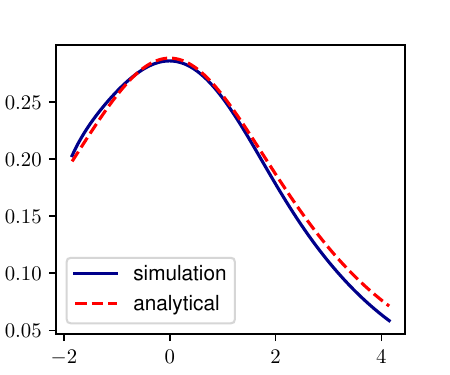}};

	\node [text width=0.01cm,align=center] at (1.3,-3.1){$N$};
 \node [text width=0.01cm,align=center] at (-6.2,-3.1){$N$};

	\node [rotate=0,text width=1cm,align=center] at (-2.7-7.5,0){$\frac{\langle\dot\sigma\rangle}{f\sqrt{\bar V}}$};
	
\node [rotate=0,text width=1cm,align=center] at (-2.2,0){$\langle\xi\rangle$};
	\end{tikzpicture}
	
	\caption{Evolution of the average value of $\dot\sigma$ (left) and $\xi$ (right) from the lattice simulation, compared with the analytical estimate of \cref{eq:xi}.  }
	\label{fig:xi}
\end{figure}
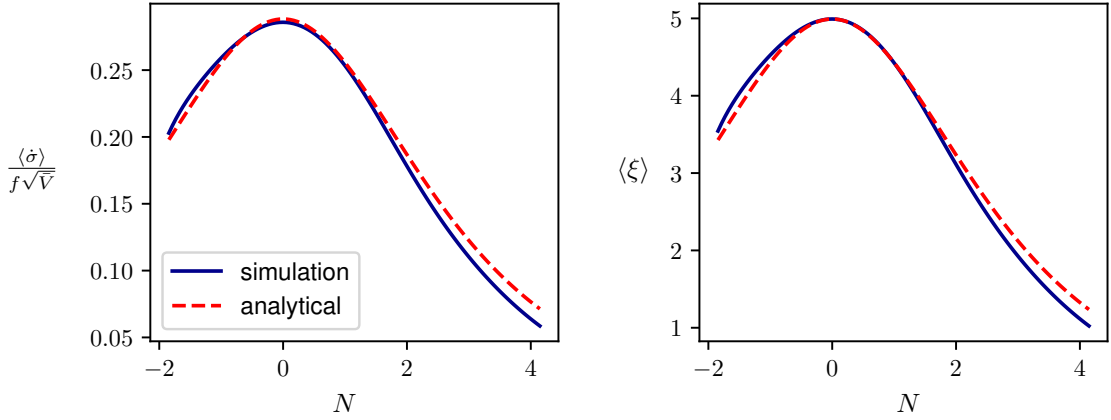

We begin by presenting the evolution of background quantities, which are obtained from the simulation as averages over the $N^3_{\rm pts}$ lattice points.
In \cref{fig:xi} we illustrate the time evolution of the background quantities $\langle\dot\sigma\rangle$ and $\langle\xi\rangle$, and compare them with the de Sitter approximation given by \cref{eq:xi}, that disregards both backreaction of the gauge fields and corrections beyond the slow-roll approximation.~\footnote{In comparing numerical vs. analytic results, we set $N=0$ when both the numerical and analytical values of $\dot\sigma$ are at their maximum.} Although the discrepancy between the lattice results and the analytical approximation is small, the axion-gauge dynamics are exponentially sensitive to slight variations in $\xi$. In the following section, we will show how this difference impacts the tachyonic amplification of the gauge field.

\subsection{Gauge field enhancement}
\label{sec:gauge_weak}
\begin{figure}
	\centering
	
	\begin{tikzpicture}
	\node (img) at (1,0) {\includegraphics[width=12cm]{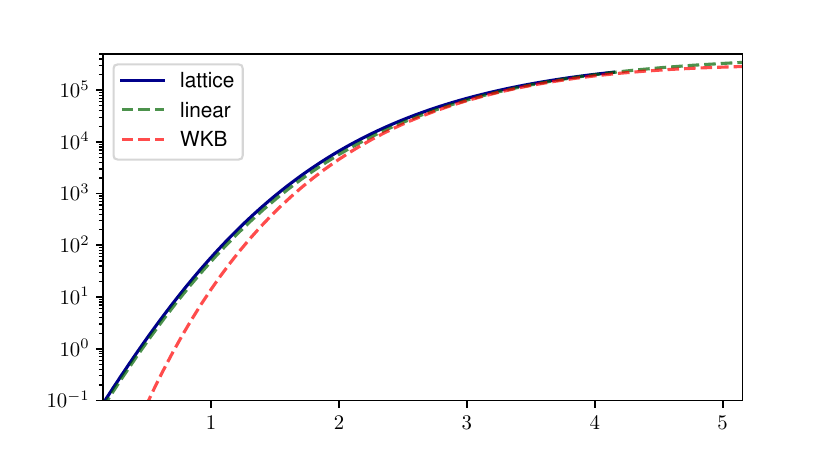}};
	
	\node [rotate=0,text width=1cm,align=center] at (-5,0){$\rho_+(\tilde k)$};
	\node [text width=0.01cm,align=center] at (1.,-3.5){$N$};

	\end{tikzpicture}
	\caption{Evolution of $\rho_+(k,\tau)$, as defined in \cref{eq:Eplus}, for a representative mode $\tilde{k} = 4 k_*$ in the case of weak backreaction. The lattice solution (blue line) is compared with the WKB solution (green dashed line) and the linear solution (red dashed line).}
	\label{fig:modeE}
\end{figure}

In this subsection, we present the results for the growth of the gauge field on the lattice, focusing on the enhanced polarization $A_{+}$. Specifically, we compute the following comoving energy density~\footnote{See Refs.~\cite{Caravano:2021pgc,Caravano:2022yyv} for details on how we compute power spectra on the lattice.}
\begin{equation}
\label{eq:Eplus}
\rho_+(k,\tau)=\langle|A^{\prime}_+(k,\tau)|^2\rangle+k^2\langle|A_+(k,\tau)|^2\rangle,
\end{equation}
and compare it to the WKB solution given by \cref{eq:WKB}, as well as to the linear solution obtained by numerically solving \cref{eq:Alin}. In \cref{fig:modeE}, we illustrate the time evolution of $\rho_+(k,\tau)$ for a characteristic mode $\tilde k \simeq 4k_*$. From this plot, we can see that the lattice correctly reproduces the linear prediction, with small differences due to the different evolution of $\xi$ (see the previous section). 

\begin{figure}
	\centering
	
	\begin{tikzpicture}
	\node (img) at (-6,0) {\includegraphics[width=7cm]{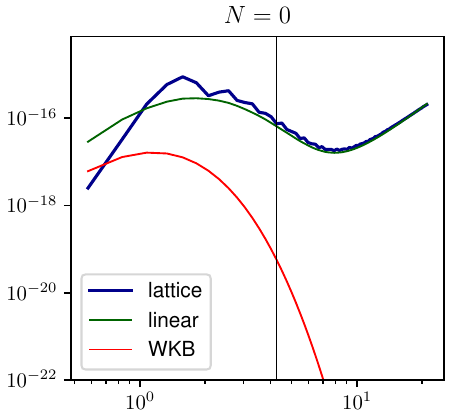}};
	
	 \node (img2) at (1.2,0) {\includegraphics[width=7cm]{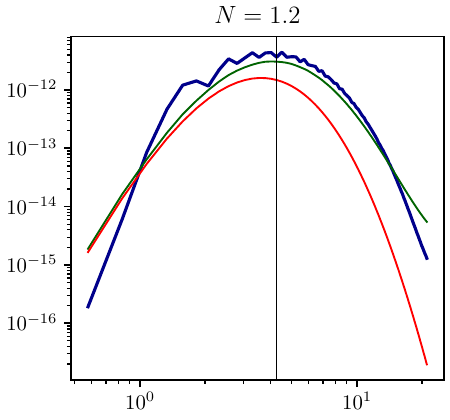}};

  \node [rotate=0,text width=1cm,align=center] at (-2.5-7.7,0){$ \frac{\rho_+(k) k^3}{2\pi^2}$};

\node [text width=0.01cm,align=center] at (1.-6.8,-3.5){$k/k_*$};
\node [text width=0.01cm,align=center] at (1.2+.2,-3.5){$k/k_*$};

\node (img) at (-6,0-7.2) {\includegraphics[width=7cm]{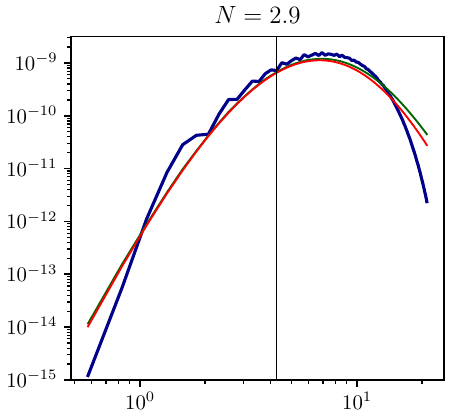}};
	
	 \node (img2) at (1.2,0-7.2) {\includegraphics[width=7cm]{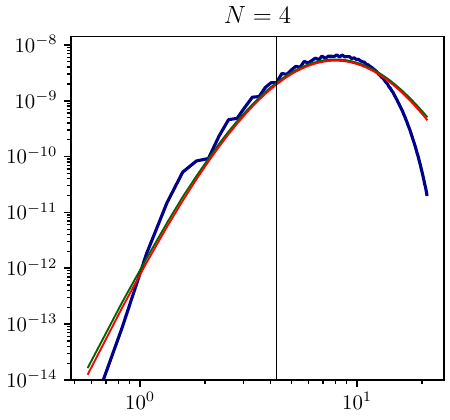}};

  \node [rotate=0,text width=1cm,align=center] at (-2.5-7.7,-7.2){$\frac{\rho_+(k) k^3}{2\pi^2}$};

\node [text width=0.01cm,align=center] at (1.-6.8,-3.5-7.2){$k/k_*$};
\node [text width=0.01cm,align=center] at (1.2+.2,-3.5-7.2){$k/k_*$};

	\end{tikzpicture}
	
	\caption{Plot of the excited gauge field polarization $\rho_+$, as defined in \cref{eq:Eplus}, at four different times during the simulation in the case of weak backreaction. The lattice results are compared with linear theory and the WKB approximation. The vertical line indicates $\tilde{k} \simeq 4 k_*$, the representative mode shown in \cref{fig:modeE}.}
	\label{fig:gauge}
\end{figure}
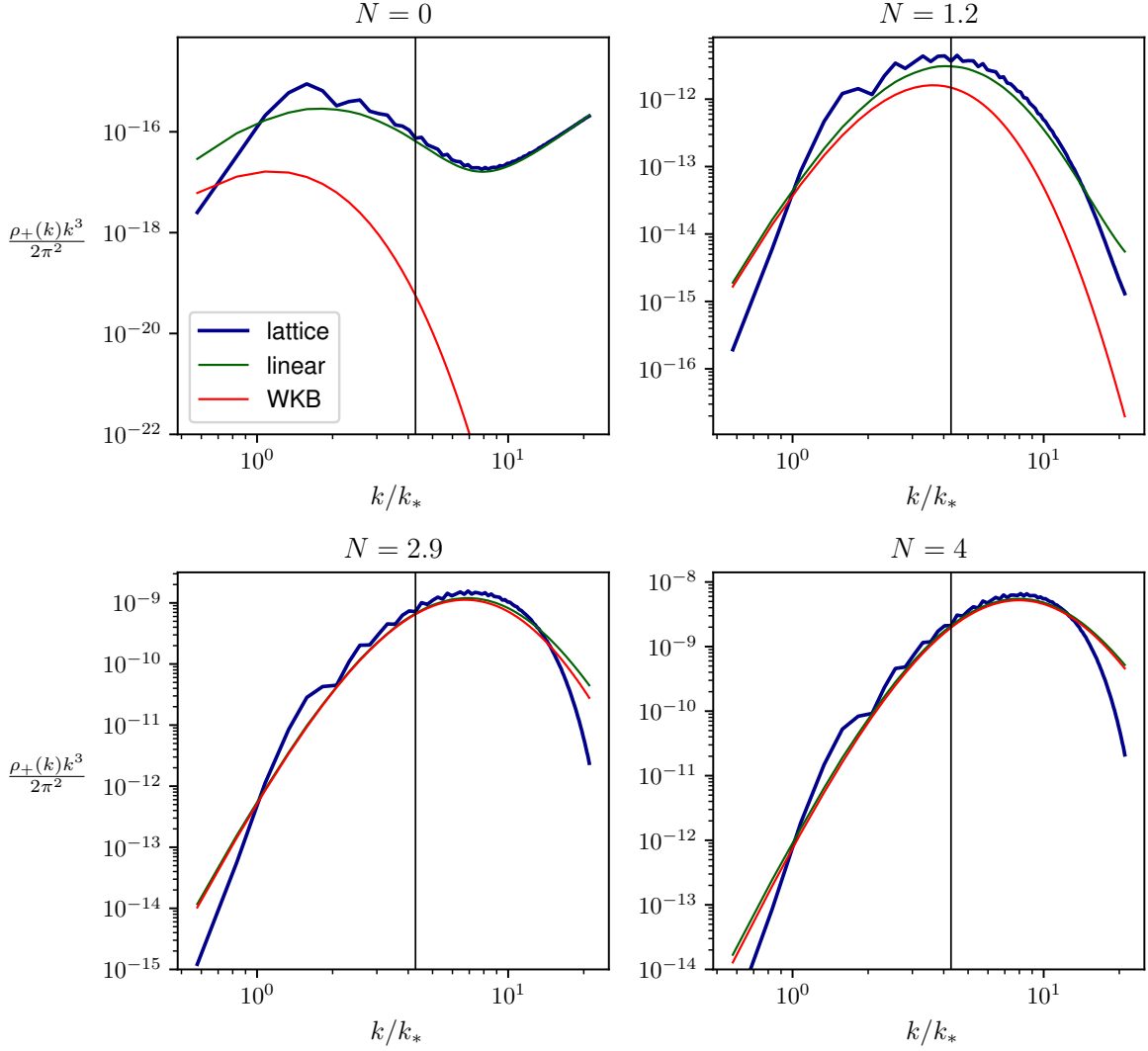

\begin{figure}
	\centering
	
	\begin{tikzpicture}
	\node (img) at (1,0) {\includegraphics[width=13cm]{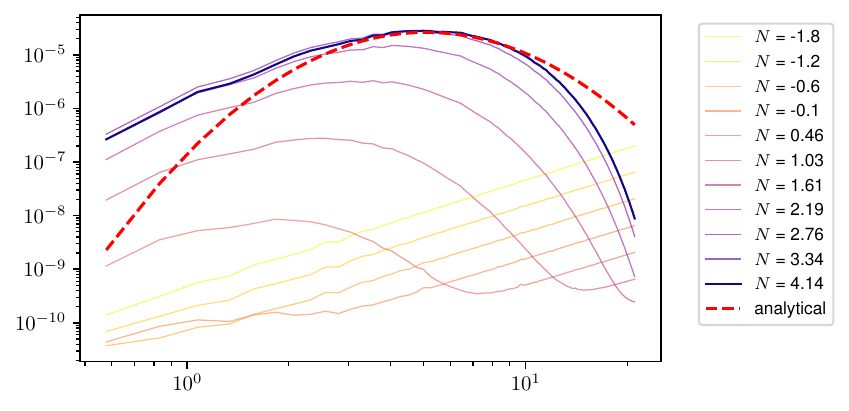}};
	
	\node [rotate=0,text width=1cm,align=center] at (-6.2,0){$\mathcal{P}_{\sigma/f}(k)$};
	\node [text width=0.01cm,align=center] at (0.,-3.3){$k/k_*$};

	\end{tikzpicture}
	
	\caption{Power spectrum of the axion field $\sigma$ in the case of weak backreaction. Different colors correspond to different times during the lattice simulation, as indicated by the legend. The final simulation result is compared with the analytical prediction from the WKB approximation (red dashed line), that we computed in \cref{app:sigma}.}
	\label{fig:ps_axion}
\end{figure}

In \cref{fig:gauge}, we show the full spectrum of $\rho_{+}(k)$ at four distinct times during the lattice simulation. We compare these results to the predictions from linear theory and the WKB solution for all modes. The discrepancy between the WKB and the numerical solution visible in the last two figures is due to the fact that the former is valid only after significant amplification of the gauge field has occurred~\cite{Namba:2015gja}. In addition, our findings indicate that, compared to linear theory, the gauge field exhibits less enhancement at large $ k $. This is consistent with the right panel of \cref{fig:xi}, which shows that $\xi$ is smaller at later times compared to the analytical estimate.

\subsection{Axion perturbation}
\label{sec:sigma_weak}
\subsubsection*{Power spectrum}
In \cref{fig:ps_axion}, we show the evolution of the power spectrum of the axion field, introduced in Fourier space as~\footnote{Details on how we compute power spectra from the lattice simulation can be found in Refs.~\cite{Caravano:2021pgc, Caravano:2022yyv}.}
\begin{equation}
\left\langle \frac{\sigma(\vec k)}{f} \, \frac{\sigma(\vec k^\prime)}{f} \right\rangle=\frac{2\pi^2}{k^3}\mathcal{P}_{\sigma/f} (k)\delta_D(\vec k +\vec k ^\prime) \;,
\end{equation}
where $\delta_D$ is the Dirac delta and $k=|\vec{k}|$.

The $\propto k^2$ profile visible at the earlier times corresponds to the Bunch-Davies vacuum initial conditions~(\ref{eq:BD}). The axion field is then enhanced due to the exponential growth of the gauge field, with the most significant growth occurring between $N = 0 $ and $ N = 2 $. The final profile has an enhancement (a ``bump'') corresponding to the scales that left the horizon during the simulation, namely while the axion had a significant roll. We compare the final time lattice result with the analytical estimate for the power spectrum, depicted as a dashed red line. We observe that the analytical estimate correctly predicts the amplitude of the power spectrum. However, the analytical solution predicts a different scale dependence. This discrepancy arises because the analytical calculation neglects the time dependence of $ H $, resulting in a different time dependence for $\xi$. Specifically, scalar modes enhanced at the beginning of the simulation (small $ k $) are expected to have a larger amplitude due to the higher value of $\xi$ (see \cref{fig:xi}). Conversely, modes excited later (large $k$) should exhibit a smaller amplitude, corresponding to the lower value of $\xi$.

\subsubsection*{Non-Gaussianity}

\begin{figure}
	\centering
	
	\begin{tikzpicture}
	\node (img) at (0,0) {\includegraphics[width=3.5cm]{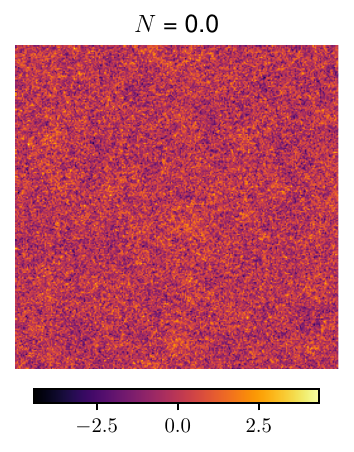}};
 \node (img) at (3.5,0) {\includegraphics[width=3.5cm]{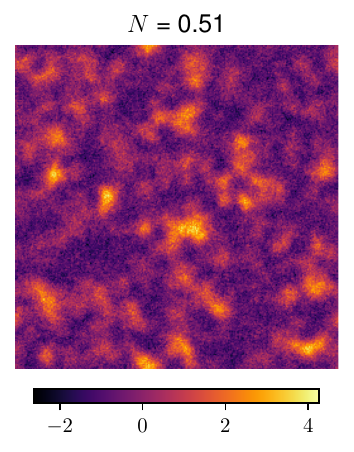}};
  \node (img) at (3.5+3.5,0) {\includegraphics[width=3.5cm]{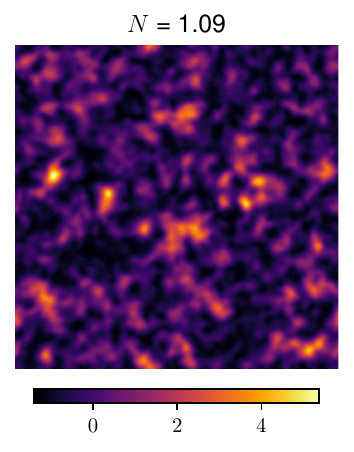}};
  
    \node (img) at (3.5+3.5+3.5,0) {\includegraphics[width=3.5cm]{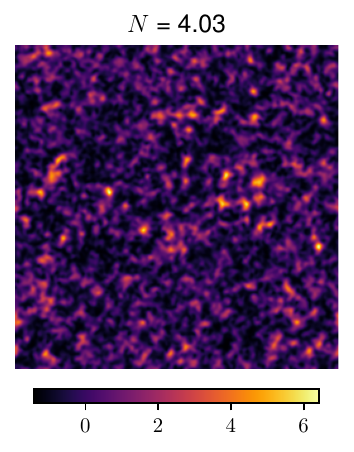}};

	\node [text width=0.01cm,align=center] at (-.6,-2.4){\footnotesize $\delta\sigma/\sqrt{\langle\delta\sigma^2\rangle}$};

	\end{tikzpicture}
	
	\caption{2D snapshot of the 3D lattice simulation in the case of negligble backreaction. The snapshots display the values of axion fluctuations $\delta\sigma=\sigma-\langle\sigma\rangle$ normalized by their standard deviation. The four panels correspond to different $e$-folding times $N$.}
	\label{fig:snap_sigma_weak}
\end{figure}

\begin{figure}
	\centering
	
	\begin{tikzpicture}
	\node (img) at (1,0) {\includegraphics[width=6.7cm]{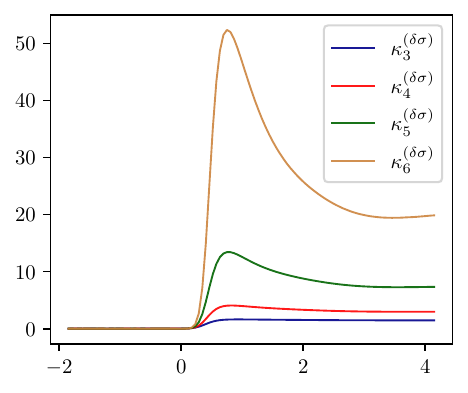}};

	\node [text width=0.01cm,align=center] at (1.2,-3.2){$N$};

 \node (img2) at (-6.5,0) {\includegraphics[width=7cm]{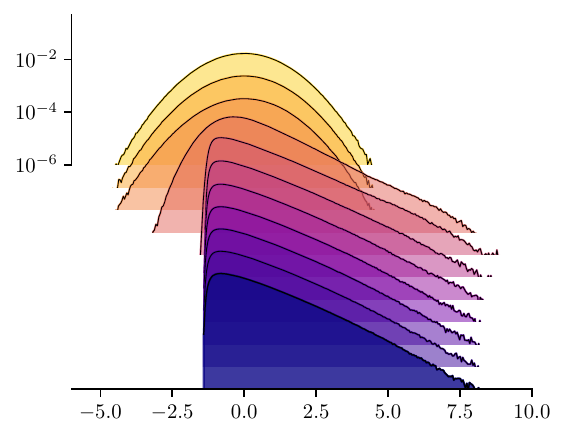}};

 \node (img2) at (-6.,3.2) {\includegraphics[width=5cm]{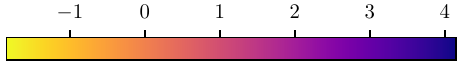}};

 \node [text width=0.01cm,align=center] at (-6,3.8){\small $N$};

	\node [text width=0.01cm,align=center] at (-6.8,-3.2){$\delta\sigma/\sqrt{\langle\delta\sigma^2\rangle}$};

	\end{tikzpicture}
	
	\caption{Non-Gaussianity of $\sigma$ in the case of weak backreaction. \textit{Left panel:} Ridgeline plot showing the normalized one-point PDF of $\delta\sigma = \sigma - \langle \sigma \rangle$ at various simulation times, indicated by the color bar. The vertical axis, representing the probability (namely, the fraction of grid points having the value indicated in the horizontal axis), is shown only for the first PDF to enhance clarity, while it is omitted for the other ones.
   \textit{Right panel:} Evolution of the cumulants defined in \cref{eq:cumulants}. }
	\label{fig:nongauss_sigma_weak}
\end{figure}
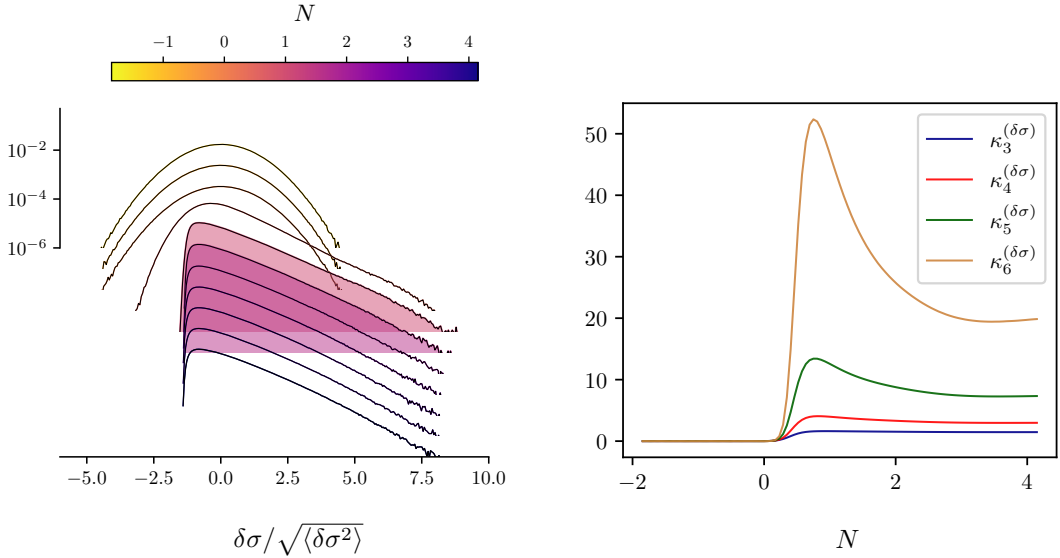

The simulation allows us to access the real-space field distribution of the axion, as shown in \cref{fig:snap_sigma_weak}. These simulation boxes can be used to compute the non-Gaussianity of scalar statistics. In the left panel of \cref{fig:nongauss_sigma_weak}
, we present the 1-point probability density function (PDF) of the axion field at different times during the simulation. The distributions are calculated as normalized histograms of the field values across the $N_{\rm pts}^3$ lattice points. We observe that the field distribution deviates significantly from Gaussian statistics. In particular, we notice an exponential tail in the distribution. This behavior is similar to what occurs in the most minimal version of this model, where the inflaton coincides with the axion~\cite{Caravano:2022epk}.

To quantify the deviation from Gaussian statistics, we introduce the following dimensionless cumulants~\cite{Bernardeau:2001qr}:
\begin{align}
		\label{eq:cumulants}
  \begin{split}
		&\kappa^{(\delta\sigma)}_3 \equiv \frac{\langle \delta\sigma^3 \rangle}{\langle\delta\sigma^2\rangle^{3/2}},\quad \kappa^{(\delta\sigma)}_4\equiv\frac{\langle \delta\sigma^4 \rangle-3\langle\delta\sigma^2\rangle^2}{\langle\delta\sigma^2\rangle^2},\quad \kappa^{(\delta\sigma)}_5\equiv\frac{\langle \delta\sigma^5 \rangle-10 \langle \delta\sigma^3 \rangle\langle\delta\sigma^2\rangle}{\langle\delta\sigma^2\rangle ^{5/2}}, \\ &\kappa^{(\delta\sigma)}_6 \equiv \frac{\langle \delta\sigma^6\rangle-10\langle \delta\sigma^3\rangle^2+30\langle\delta\sigma^2\rangle^3-15\langle\delta\sigma^4\rangle\langle\delta\sigma^2\rangle}{\langle\delta\sigma^2\rangle^3},
  \end{split}
	\end{align}
that all vanish in the Gaussian case. In the right panel of \cref{fig:nongauss_sigma_weak}, we show the evolution of these cumulants during the simulation. We can see that higher-order cumulants are very large and do not converge. This behavior reflects the presence of the exponential tail in the PDF, and it is analogous to what is reported in Ref.~\cite{Caravano:2022epk} for the minimal model in which the axion is the inflaton. Note that this implies that the bispectrum contains very limited information about the non-Gaussianity of the axion field. 

The exponential tail of the final-time PDF of $\delta\sigma$ can be fitted by the function $ f(x) = e^{-m x} $, where $ m \simeq 1.2 $. It would be interesting to investigate how the quantity $ m $ depends on theoretical parameters such as $ \xi_* $ and $ \delta $. We leave this exploration for future work. Note that the tail of the PDF is different from what we would expect from a $\chi^2$ distribution (which is the simplest guess that is reasonable to make for the PDF that results from the convolution of two gaussian gauge modes~\cite{Linde:2012bt}), which also has an exponential behaviour in the tail, but with $ f(x) \propto e^{- x/2}/\sqrt{x}$, for $x\gg 1$~\cite{Linde:2012bt}.

\subsection{Curvature perturbation}

\begin{figure}
	\centering
	
	\begin{tikzpicture}
	\node (img) at (1,0) {\includegraphics[width=13cm]{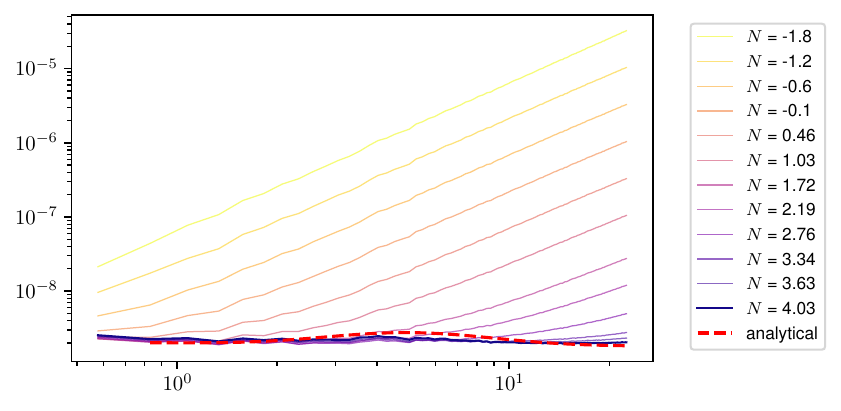}};
	
	\node [rotate=0,text width=1cm,align=center] at (-6.2,0){$\mathcal{P}_{\zeta}(k)$};
	\node [text width=0.01cm,align=center] at (0.,-3.3){$k/k_*$};

	\end{tikzpicture}
	
	\caption{Power spectrum of $\zeta$ in the case of weak backreaction. At the final time, we compare the lattice result with the analytical prediction computed in \cite{Namba:2015gja}, depicted as a red dashed line.}
	\label{fig:pzeta_weak}
\end{figure}

With the choice of parameters adopted in this section, the curvature perturbation $\zeta$ sourced by the axion field is of the same order as the standard vacuum contribution, according to the analytical estimate of Ref.~\cite{Namba:2015gja}. More precisely, the sourced contribution estimated analytically is approximately 40\% of the vacuum one at the peak scale. This is consistent with the simulation results, as shown in \cref{fig:pzeta_weak}. The plot shows that the power spectrum from the simulation remains close to the nearly scale-invariant prediction with amplitude $\mathcal P_{\zeta, \rm vacuum}\simeq 2 \times 10^{-9}$. The power spectrum shows only a small peak of about 20\% at the peak scale, roughly consistent with the analytical estimate. The comoving curvature perturbation is computed from the lattice using the linear relation $\zeta = -\delta\phi H/\dot\phi$.

\subsection{Energy contributions}
\label{sec:energy_weak}
\begin{figure}
	\centering
	
	\begin{tikzpicture}
	\node (img) at (1,0) {\includegraphics[width=13cm]{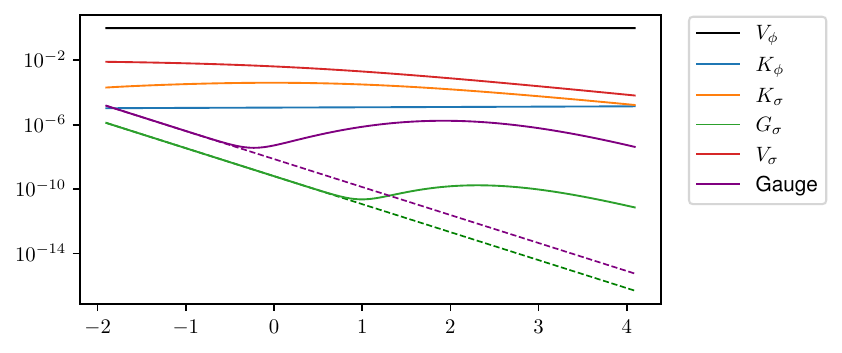}};

	\node [text width=0.01cm,align=center] at (0.,-2.9){$N$};
	\end{tikzpicture}
	
	\caption{Evolution of different components contributing to the total energy density. For a field $f$, $K_f$ and $G_f$ represent its kinetic and gradient energy, respectively, while $V_f$ denotes its potential energy. The purple curve illustrates the total energy density of the gauge field. These quantities are computed as average energies over the lattice simulation. The dashed lines represent the vacuum contributions, as explained in \cref{sec:energy_weak}.} 
	\label{fig:energy}
\end{figure}

Before proceeding to the strong backreaction case, we present the evolution of the different contributions to the mean energy density $\langle \rho \rangle$. In \cref{fig:energy}, we plot the evolution of all contributions during the simulation. We observe that the inflaton potential remains the dominant component. Additionally, the kinetic energy of the axion, $K_\sigma$, remains much larger than the energy density of the gauge field. This is expected in the regime with small backreaction. In this figure, the dotted purple and green lines represent the vacuum contributions to the gauge field energy density and the gradient energy of the axion field, respectively. These vacuum contributions diverge in the limit of infinite UV resolution and should be subtracted. However, they play a negligible role in the evolution of the system. This is confirmed in \cref{app:resolution}, where we show that changing the UV cutoff, which corresponds to changing the amount of vacuum energy, does not affect any of our results. 

Finally, we note that once the gauge field is enhanced, it does not immediately source axion perturbations. Axion particle production occurs roughly $1.5$ $e$-folds later. This delay is crucial for understanding how backreaction operates in this system, as discussed in \cref{sec:back_strong}.

\section{Strong backreaction}
\label{sec:results_strong}

We now turn to the case of strong backreaction. To do this, we maintain the same parameters as in the previous section and increase the axion-gauge coupling to $\alpha = 24.17$, corresponding to a maximum value $\xi = 6$ (according to the background simulation, see the beginning of \cref{sec:weak}). This is the value required to explain the PTA signal, according to the analytical calculations that neglect backreaction~\cite{Unal:2023srk}. In this regime, however, we expect significant deviations from the linear theory and the WKB approximation~\cite{Peloso:2016gqs,Campeti:2022acx,Unal:2023srk}.

\subsection{Background evolution}
\label{sec:back_strong}
\begin{figure}
	\centering
	
	\begin{tikzpicture}
		\node (img) at (1.3,0) {\includegraphics[width=7.cm]{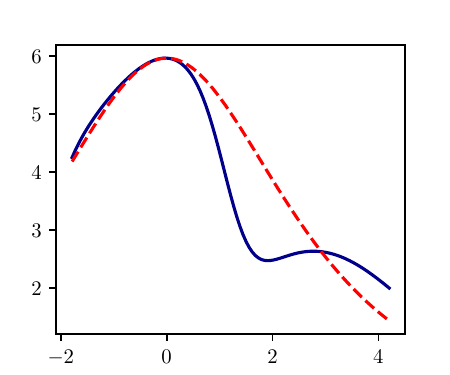}};
	\node (img2) at (-6.,0) {\includegraphics[width=7.cm]{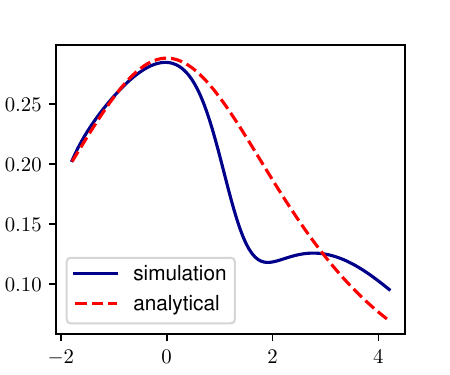}};

	\node [text width=0.01cm,align=center] at (1.3,-3.1){$N$};
	\node [text width=0.01cm,align=center] at (-6.2,-3.1){$N$};
	
	\node [rotate=0,text width=1cm,align=center] at (-2.7-7.5,0){$\frac{\langle\dot\sigma\rangle}{f\sqrt{\bar V}}$};
	
	\node [rotate=0,text width=1cm,align=center] at (-2.2,0){$\langle\xi\rangle$};
	\end{tikzpicture}
	
	\caption{Evolution of $\dot\sigma$ (left panel) and $\xi$ (right panel) in the case of strong backreaction, compared with the analytical estimate of \cref{eq:xi}. }
	\label{fig:xi_strong}
\end{figure}
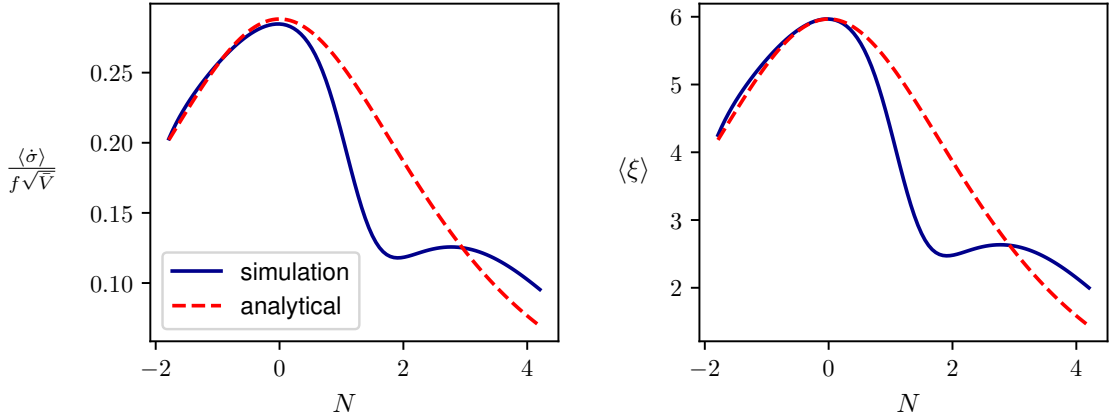

In \cref{fig:xi_strong} we show the evolution of $\dot \sigma$ and $\xi$ during the lattice simulation and compare it to the linear theory prediction. The background dynamics is clearly affected by the gauge field growth. The deviation manifests itself as an oscillatory feature about the background solution with a half-period of $\Delta N\simeq 1.5$ e-folds. This oscillation is caused by the fact that, when the gauge field is enhanced, it sources scalar perturbations with a delay, as we discussed already at the end of \cref{sec:energy_weak}. This behavior is similar to what happens in the minimal axion-U(1) model of inflation, where the axion is the inflaton \cite{Cheng:2015oqa,Notari:2016npn,DallAgata:2019yrr,domcke2020resonant,Caravano:2022epk,Figueroa:2023oxc,Peloso:2022ovc}, explained in \cref{sec:introduction} (see \cref{fig:illustration}).
\subsection{Gauge field enhancement}
\begin{figure}
	\centering
	
	\begin{tikzpicture}
		\node (img) at (1,0) {\includegraphics[width=12cm]{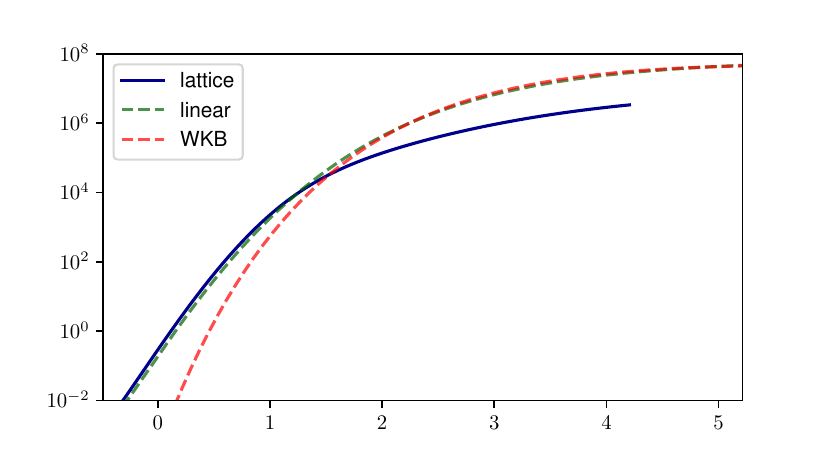}};
		
		\node [rotate=0,text width=1cm,align=center] at (-5,0){$\rho_+(\tilde k)$};
		\node [text width=0.01cm,align=center] at (1.,-3.5){$N$};

	\end{tikzpicture}
	
	\caption{Same as \cref{fig:modeE}, but in the case of strong backreaction. In this case, the lattice solution (blue) is suppressed compared to the WKB solution (green dashed) and to the linear solution (red dashed line).}
	\label{fig:modeE_strong}
\end{figure}

\begin{figure}
	\centering
		\begin{tikzpicture}
		\node (img) at (-6,0) {\includegraphics[width=7cm]{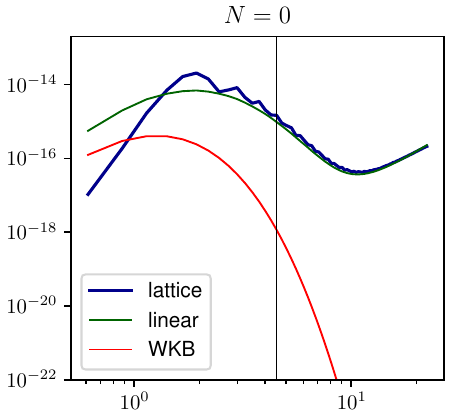}};
		
		\node (img2) at (1.2,0) {\includegraphics[width=7cm]{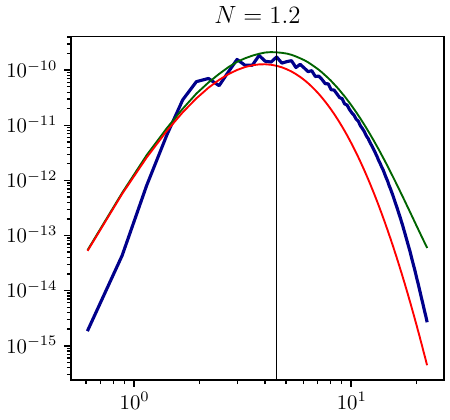}};
		
		\node [rotate=0,text width=1cm,align=center] at (-2.5-7.7,0){$ \frac{\rho_+(k) k^3}{2\pi^2}$};

		\node [text width=0.01cm,align=center] at (1.-6.8,-3.5){$k/k_*$};
		\node [text width=0.01cm,align=center] at (1.2+.2,-3.5){$k/k_*$};

		\node (img) at (-6,0-7.2) {\includegraphics[width=7cm]{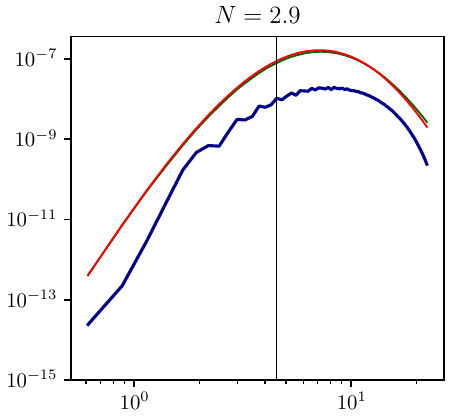}};
		
		\node (img2) at (1.2,0-7.2) {\includegraphics[width=7cm]{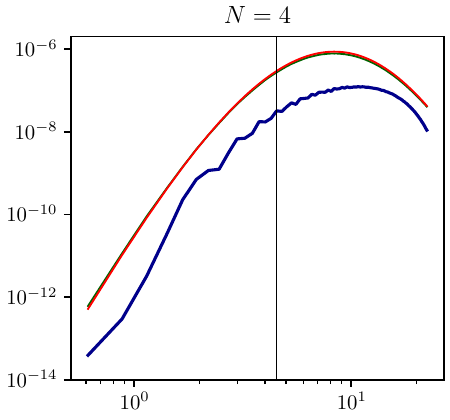}};
		
		\node [rotate=0,text width=1cm,align=center] at (-2.5-7.7,-7.2){$ \frac{\rho_+(k) k^3}{2\pi^2}$};

		\node [text width=0.01cm,align=center] at (1.-6.8,-3.5-7.2){$k/k_*$};
		\node [text width=0.01cm,align=center] at (1.2+.2,-3.5-7.2){$k/k_*$};

	\end{tikzpicture}
	
	\caption{Same as \cref{fig:gauge}, but in the case of strong backreaction. The vertical line shows $\tilde k \simeq 4k_*$, the representative mode depicted in \cref{fig:modeE_strong}.}
	\label{fig:gauge_strong}
\end{figure}
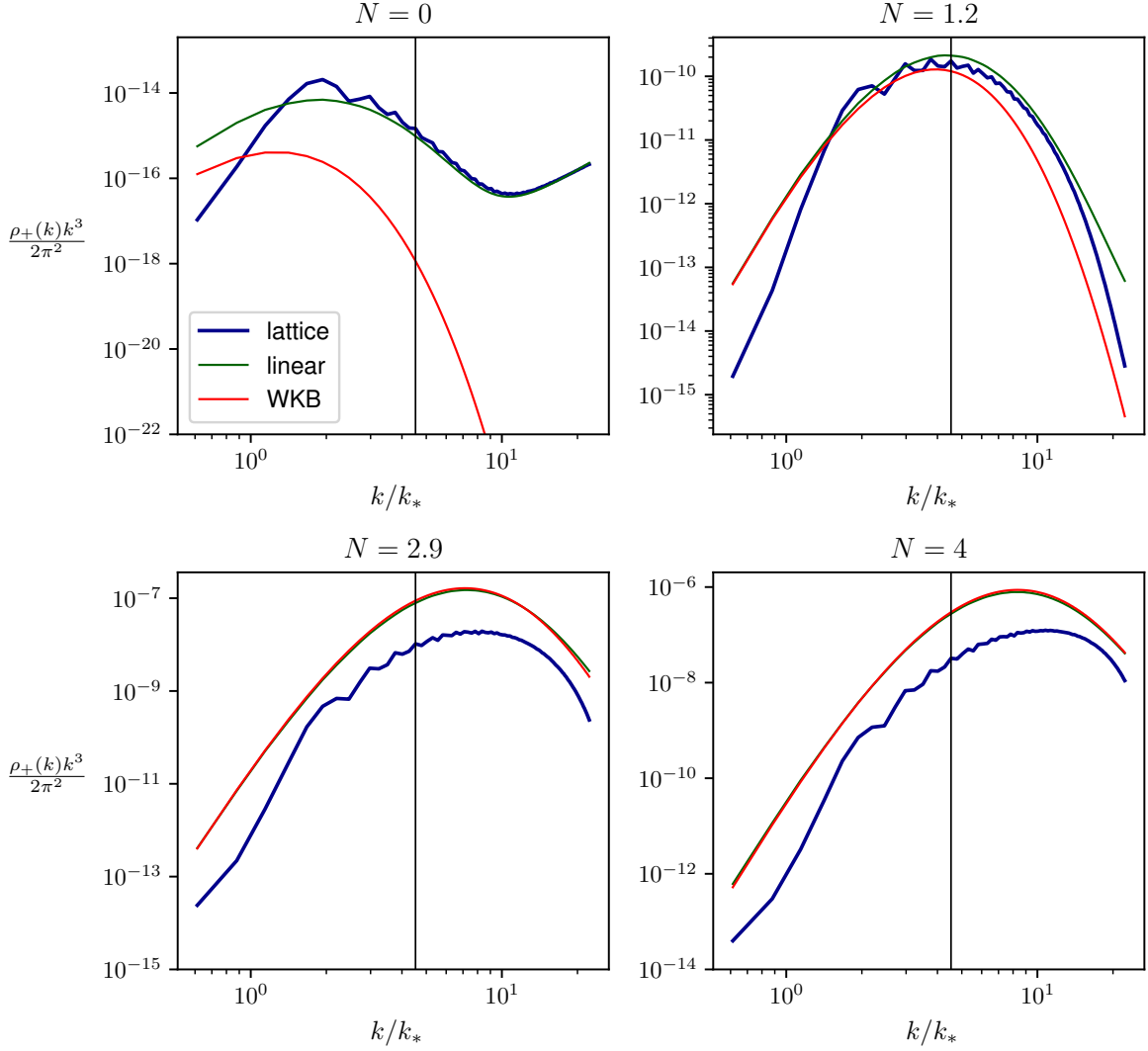

In \cref{fig:modeE_strong} we show the time evolution of $\rho_+$, defined in \cref{eq:Eplus}, for one characteristic mode $\tilde k \simeq 4k_*$. From this plot, we can see that the lattice solution deviates from the analytical and linear estimates, giving a much lower value of $\rho_+(\tilde k)$ for $N>2$. In \cref{fig:gauge_strong}, we plot the power spectrum of $\rho_+$ at  different times during the simulation. These plots show that the WKB approximation matches well the linear computation at late times. However, both approximations overestimate the amplitude of the power spectrum resulting from the lattice dynamics at times when the gauge field is excited the most. This is due to the fact that now the backraction of the gauge fields and the axion dynamics (not included in the WKB and in the linear computaton) is significant. 

These results show that the gauge field growth is significantly suppressed by the backreaction. This is expected because, as we have seen in the last section, the growth of the gauge field dampens the velocity of the axion, resulting in a lower value of $\xi$ during the evolution (see \cref{fig:xi_strong}).

\subsection{Axion perturbation}

\subsubsection*{Power spectrum}
\begin{figure}
	\centering
	
	\begin{tikzpicture}
	\node (img) at (1,0) {\includegraphics[width=13cm]{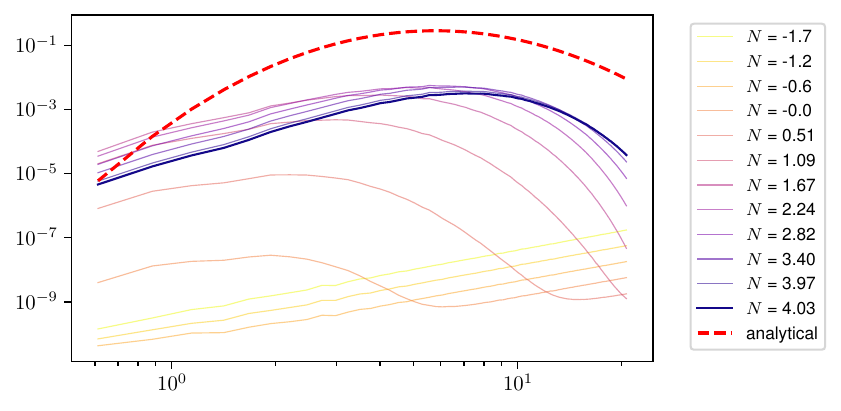}};
	
	\node [rotate=0,text width=1cm,align=center] at (-6.2,0){$\mathcal{P}_{\sigma/f}(k)$};
	\node [text width=0.01cm,align=center] at (0.,-3.3){$k/k_*$};

	\end{tikzpicture}
	
	\caption{Power spectrum of the axion field $\sigma$ in the scenario of strong backreaction, depicted at various times during the lattice simulation. The final simulation result (blue) is compared with the prediction from the WKB approximation (red dashed line).}
	\label{fig:ps_axion_strong}
\end{figure}

As a result of suppressed gauge field growth, backreaction also suppresses the power spectrum of the axion field. This effect is illustrated in \cref{fig:ps_axion_strong}, which displays the power spectrum of $\sigma$ computed from the lattice simulation, compared to the analytical prediction at the final time. Note that, even if we are in the strong backreaction regime, the axion power spectrum is still perturbative, $\mathcal P_{\sigma/f}\ll 1$. Therefore, we can still trust our hybrid numerical approach in the calculation of $\delta\phi$, which will be computed in \cref{sec:curvature_strong}. 

\subsubsection*{Non-Gaussianity}

\begin{figure}
	\centering
	
	\begin{tikzpicture}
	\node (img) at (0,0) {\includegraphics[width=3.5cm]{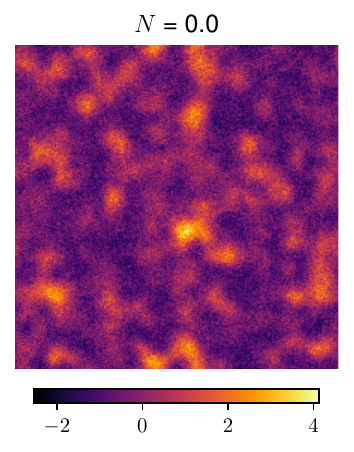}};
 \node (img) at (3.5,0) {\includegraphics[width=3.5cm]{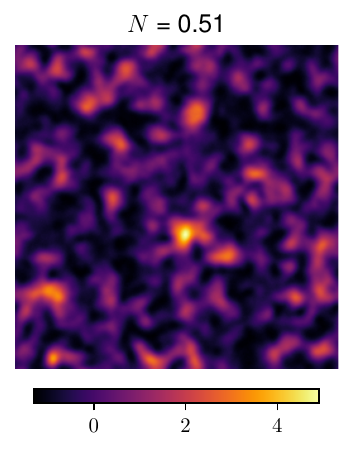}};
  \node (img) at (3.5+3.5,0) {\includegraphics[width=3.5cm]{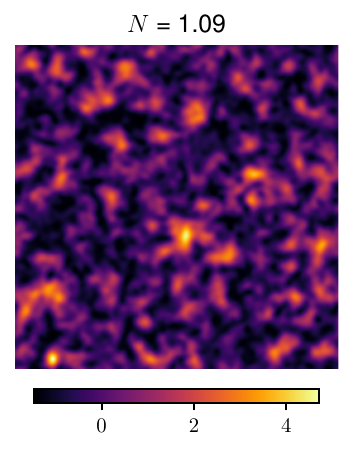}};
    \node (img) at (3.5+3.5+3.5,0) {\includegraphics[width=3.5cm]{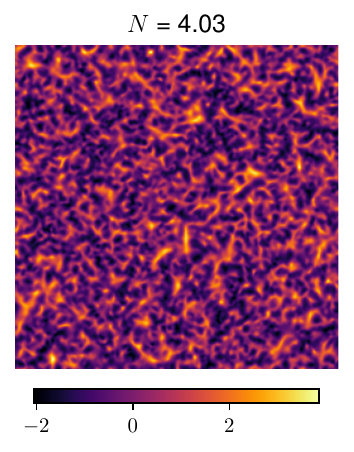}};

	\node [text width=0.01cm,align=center] at (-.6,-2.4){\footnotesize $\delta\sigma/\sqrt{\langle\delta\sigma^2\rangle}$};

	\end{tikzpicture}
	
	\caption{Same as \cref{fig:snap_sigma_weak}, but in the case of strong backreaction, showing values of the axion field $\sigma$ across a 2D slice of the simulation.}
	\label{fig:snap_sigma_strong}
\end{figure}

\begin{figure}
	\centering
	
	\begin{tikzpicture}
	\node (img) at (1,0) {\includegraphics[width=6.7cm]{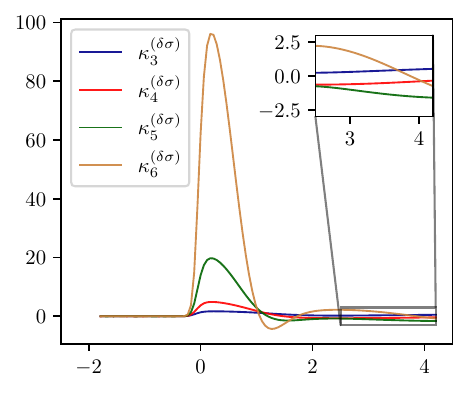}};

	\node [text width=0.01cm,align=center] at (1.2,-3.2){$N$};

 \node (img2) at (-6.5,0) {\includegraphics[width=7cm]{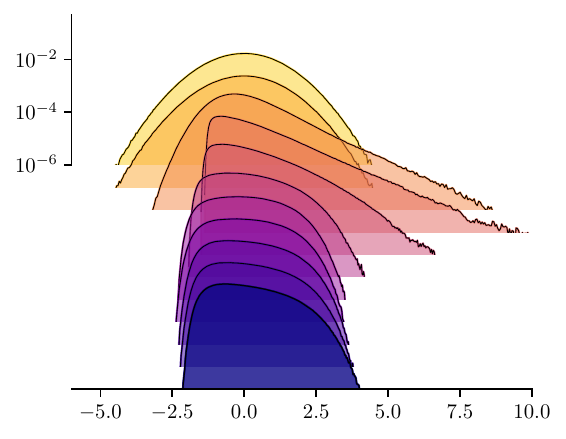}};

 \node (img2) at (-6.,3.2) {\includegraphics[width=5cm]{plots/hist_zeta_colorbar.pdf}};

 \node [text width=0.01cm,align=center] at (-6,3.8){\small $N$};

	\node [text width=0.01cm,align=center] at (-6.8,-3.2){$\delta\sigma/\sqrt{\langle\delta\sigma^2\rangle}$};

	\end{tikzpicture}
	
	\caption{Same as \cref{fig:nongauss_sigma_weak}, showing non-Gaussianity of $\sigma$ in the case of strong backreaction. The right panel includes an inset that provides a detailed view of the region $ N > 2.5$.}
	\label{fig:nongauss_sigma_strong}
\end{figure}

In \cref{fig:snap_sigma_strong}, we present snapshots of our simulation boxes for the axion field $\sigma$ in the case of strong backreaction. As we did in \cref{sec:sigma_weak}, we compute the 1-point PDFs of the axion field across the simulation and analyze their cumulants. The results are displayed in \cref{fig:nongauss_sigma_strong}.

We observe that non-Gaussianity is significantly suppressed compared to the case with weak backreaction. This suppression is consistent with the interpretation given in Ref.~\cite{Caravano:2022epk}, which attributes the reduction in non-Gaussianity to the large number of excited gauge field modes, leading to Gaussian statistics due to the central limit theorem. Notably, unlike the minimal model discussed in Ref.~\cite{Caravano:2022epk}, non-Gaussianity remains sizable and $O(1)$ at the end of the simulation. This is consistent with the central limit theorem interpretation, as the strong scale dependence in this model (due to the fact that the axion roll is significant only for a few e-folds) limits the number of excited gauge field modes compared to the minimal scenario.

\subsection{Curvature perturbation}
\label{sec:curvature_strong}
We now come to the results for the comoving curvature perturbation, that we compute from the lattice using the linear relation $\zeta = -\delta\phi H/\dot\phi$. 

\begin{figure}
	\centering
	
	\begin{tikzpicture}
	\node (img) at (1,0) {\includegraphics[width=13cm]{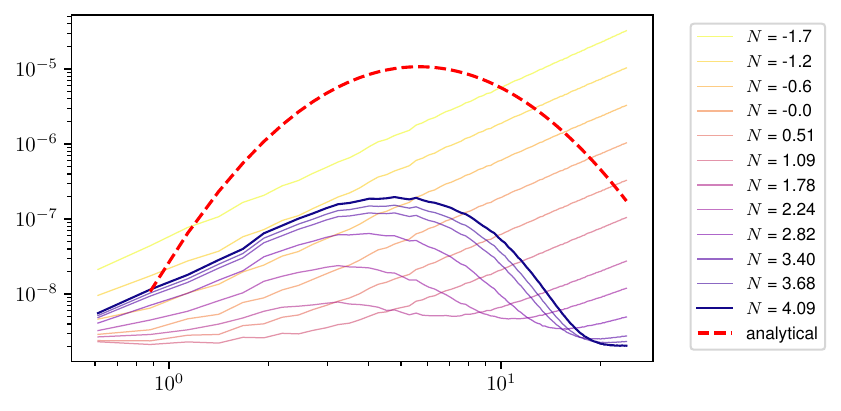}};
	
	\node [rotate=0,text width=1cm,align=center] at (-6,0){$\mathcal{P}_{\zeta}(k)$};
	\node [text width=0.01cm,align=center] at (0.,-3.2){$k/k_*$};

	\end{tikzpicture}
	
	\caption{Power spectrum of $\zeta$ in the case of strong backreaction. Different colors correspond to different simulation times, while the red dashed line shows the analytical calculation.}\label{fig:ps_zeta_strong}
\end{figure}

\subsubsection*{Power spectrum}

In \cref{fig:ps_zeta_strong}, we present the power spectrum of $\zeta$ at various times during the lattice simulation and compare it to the WKB approximation at the final time. The scalar power spectrum is significantly suppressed relative to the analytical estimate. This suppression is a consequence of the reduced gauge field growth due to strong backreaction. The sourced contribution is still much larger than the standard vacuum contribution $\mathcal{P}_{\zeta,\rm vacuum}= 2\times 10^{-9}$.
\subsubsection*{Non-Gaussianity}
\begin{figure}
	\centering
	
	\begin{tikzpicture}
	\node (img) at (0,0) {\includegraphics[width=3.5cm]{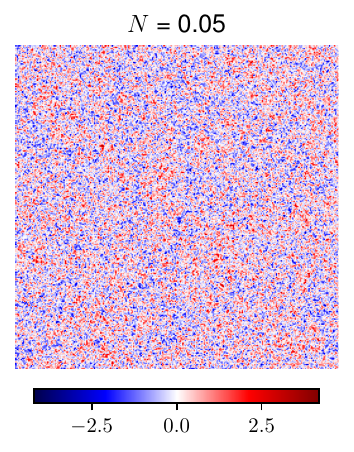}};
 \node (img) at (3.5,0) {\includegraphics[width=3.5cm]{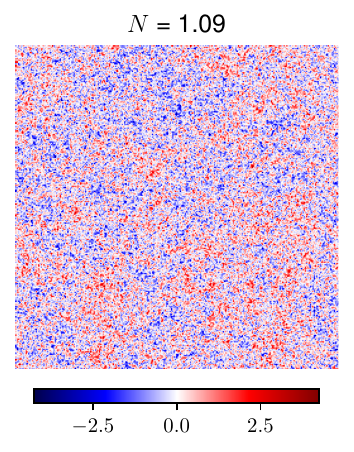}};
  \node (img) at (3.5+3.5,0) {\includegraphics[width=3.5cm]{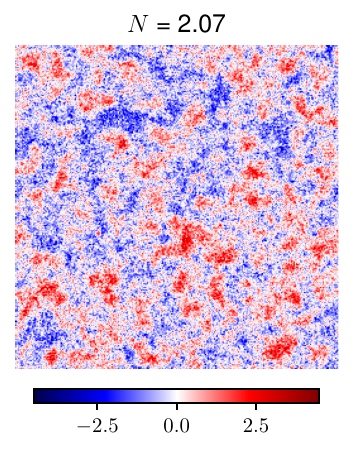}};
  
    \node (img) at (3.5+3.5+3.5,0) {\includegraphics[width=3.5cm]{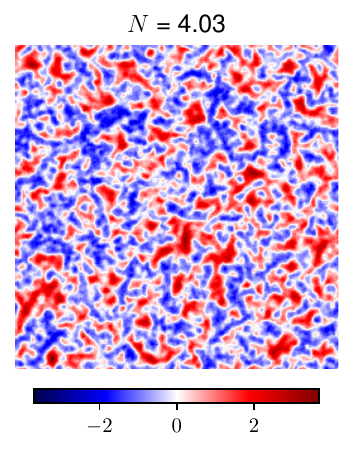}};

	\node [text width=0.01cm,align=center] at (-.6,-2.4){\footnotesize $\zeta/\sqrt{\langle\zeta^2\rangle}$};

	\end{tikzpicture}
	
	\caption{Simulation snapshots for $\zeta$ in the case of strong backreaction.}
	\label{fig:snap_zeta}
\end{figure}
\begin{figure}
	\centering
	
	\begin{tikzpicture}
	\node (img) at (1,0) {\includegraphics[width=6.7cm]{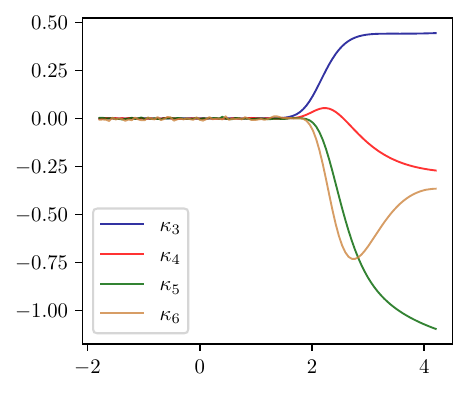}};

	\node [text width=0.01cm,align=center] at (1.2,-3.2){$N$};

 \node (img2) at (-6.5,0) {\includegraphics[width=7cm]{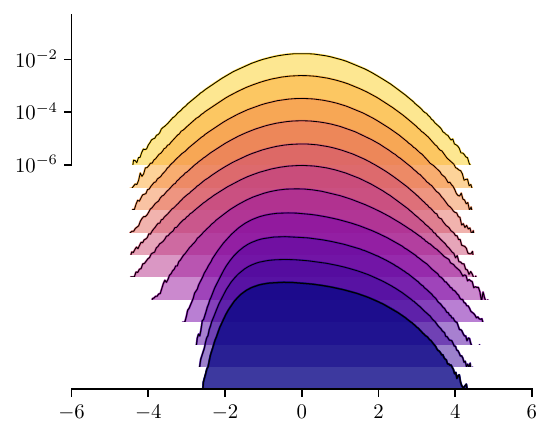}};

 \node (img2) at (-6.,3.2) {\includegraphics[width=5cm]{plots/hist_zeta_colorbar.pdf}};

 \node [text width=0.01cm,align=center] at (-6,3.8){\small $N$};

	\node [text width=0.01cm,align=center] at (-6.8,-3.2){$\zeta/\sqrt{\langle\zeta^2\rangle}$};

	\end{tikzpicture}
	
	\caption{Non-Gaussianity of $\zeta$ in the case of strong backreaction. See the caption of \cref{fig:nongauss_sigma_weak} for the explanation of the different panels.}
	\label{fig:nongauss_zeta}
\end{figure}
We now discuss the non-Gaussianity of $\zeta$. In \cref{fig:snap_zeta}, we present 2D snapshots from the simulation. From the full 3D simulation boxes, we extract the 1-point PDF of $\zeta$ and its cumulants, as shown in \cref{fig:nongauss_zeta}. Non-Gaussianity of the observable $\zeta$ on super-horizon scales is large and non-trivial. In particular, we observe that a large amount of non-Gaussianity is encoded in cumulants beyond the 3rd order. This implies that the bispectrum will carry limited information about the statistical properties of the primordial curvature fluctuation. This has crucial phenomenological implication, as we discuss in the Conclusion (\cref{sec:conclusion}).

\subsection{Energy contributions}
\begin{figure}
	\centering
	
	\begin{tikzpicture}
	\node (img) at (1,0) {\includegraphics[width=13cm]{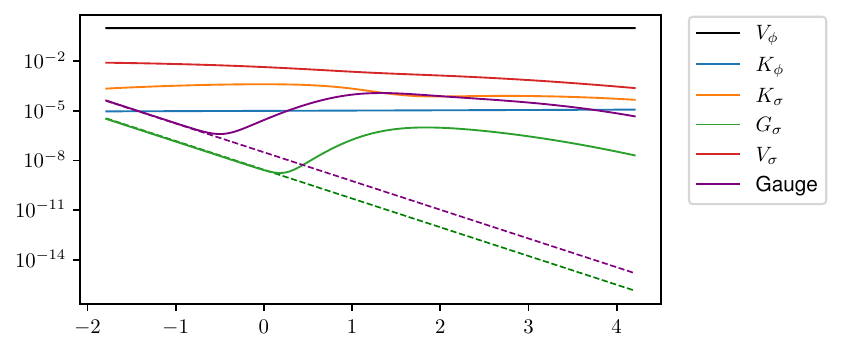}};

	\node [text width=0.01cm,align=center] at (0,-2.9){$N$};
	\end{tikzpicture}
	
	\caption{Energy contributions in the case of strong backreaction. See the caption of \cref{fig:energy} for the definition of the different components.}
	\label{fig:energy_strong}
\end{figure}

Finally, in \cref{fig:energy_strong}, we show the evolution of the various contributions to the energy density, which source the evolution of the Universe through the Friedmann equations. We can see that, even in the strong backreaction regime, the energy density of the Universe is dominated by the background inflaton field. However, in this scenario, the mean energy density of the gauge field becomes comparable to the kinetic energy of the axion field $K_\sigma$ around $N=2$. This indicates that the gauge field cannot be neglected in the background dynamics of the axion field, signaling significant backreaction in the system. A close inspection to the $K_\sigma$ profile shows the oscillatory behaviour of the axion kinetic energy at the moment in which the backreaction is more relevant, as we also observed in~\cref{fig:xi_strong}. Moreover, we observe that the gradient energy of the axion field (which is the cause of the suppression of the oscillatory behaviour observed in the simulations of~\cite{Figueroa:2023oxc} in the model in which the inflaton axion is continuously rolling) remains small in this simulation.

\section{Conclusions}
\label{sec:conclusion}
Understanding axion-gauge systems during inflation is of utmost importance due to the theoretical and observational interest in these models. In particular, it is important to develop a nonlinear and non-perturbative understanding of these systems, as the observationally relevant regime is often beyond the validity of standard perturbation theory. Lattice simulations of inflation are starting to make significant progress in this direction. In this work, we considered a fast-rolling axion coupled to a gauge field, both minimally coupled to the main inflaton sector. We performed the first numerical lattice simulation of this system, employing a novel numerical scheme.

In the regime where standard perturbation theory techniques are valid, we used the simulation to verify the accuracy of analytical approximations. While we found that the magnitude of the sourced axion fluctuations is consistent with analytical predictions, the lattice simulation provides a more precise characterization of the scale dependence of this signal. This improvement is due to our ability to relax the assumption of a constant Hubble rate $H$ and to go beyond the slow-roll approximation for the dynamics of the axion. 
More importantly, the simulations allow us to understand non-Gaussianity resulting from the nonlinear axion-gauge interactions beyond just the three-point function. We showed that scalar fluctuations exhibit a non-Gaussian exponential tail, with high-order information being crucial for describing the statistics. This finding is consistent with results from the minimal version of this model, where the axion is the inflaton and the sourced contribution does not have such a strong scale dependence. From this, we learn that the exponential tail is a general feature of this model and is not linked to the scale dependence of the signal.

After exploring the linear regime, we investigated the regime strongly affected by backreaction, where standard perturbation theory is not expected to provide an accurate picture. This regime is extremely relevant because many interesting observational consequences, such as a significant gravitational wave background at PTA scales, lie precisely in this regime. We found that backreaction strongly suppresses the growth of the gauge field. As a consequence, the amplitude of sourced scalar perturbations is also suppressed. Still, we see from Figure~\ref{fig:ps_zeta_strong} that the sourced perturbations are greater than the vacuum ones in the regime of strong backreaction. Although the case we studied has significant differences with that of an axion inflation, it is hard to imagine that a prolonged motion of the axion throughout inflation, and a direct coupling to the gauge fields of the field responsible for the density perturbations will result in a smaller sourced signal than the one emerged from our study. Therefore our results confirm that a regime of strong backreaction in axion inflation~\cite{Anber:2009ua} can be at play only well after the CMB modes have been produced. 

Concerning the statistics of the perturbations, we showed that the exponential tail in the scalar distribution is suppressed by backreaction. This is due to the large number of excited gauge field modes, which tend to `Gaussianize' the system due to the central limit theorem. However, we found that the suppression is less severe compared to the minimal model where the axion is the inflaton, due to the strong scale dependence in this model. Therefore, while the scale dependence does not affect the presence of an exponential tail, it influences the suppression of this tail in the strong backreaction regime. As a result, high-order non-Gaussianity can still play an important role in this system when backreaction is at play.

These non-perturbative non-Gaussianities have crucial observational implications. First, high-order information, neglected in previous studies, is expected to play an important role in shaping the large-scale structure of our Universe. This will have observational consequences for the statistical properties of large-scale structures and the CMB (see, for example, \cite{Munchmeyer:2019wlh,Philcox:2024jpd,Coulton:2024vot}). Moreover, the non-trivial non-Gaussian statistics will have important implications for the formation of primordial black holes on small scales, which is highly sensitive to the statistical properties of the tail of the distribution~\cite{Ezquiaga:2019ftu,Figueroa:2020jkf,Kitajima:2021fpq,Biagetti:2021eep,LISACosmologyWorkingGroup:2023njw}. 

As already mentioned, this model can contribute to the gravitational wave signal observed by PTA experiments. However, this requires working in the strong backreaction regime where analytical calculations cannot be trusted, as discussed in Ref.~\cite{Unal:2023srk}. Although we did not calculate gravitational waves directly, we showed that backreaction significantly suppresses the growth of the gauge field in this regime, which is the primary source of the gravitational wave background. Therefore, we conclude that the parameter choices outlined in Ref.~\cite{Unal:2023srk} are likely to result in a gravitational wave signal that is too weak to be detected by PTA experiments, due to the suppression effect observed in our simulations. Future dedicated work, employing higher-resolution lattice simulations, will be necessary to identify the viable parameter space for a detectable PTA signal. In this paper, we focused on the theoretical understanding of this system. All the phenomenological and observational prospects outlined in these Conclusions are left for future work.

\section*{Acknowledgements}
We are grateful to Drew Jamieson, Eiichiro Komatsu, Ippei Obata, S\'ebastien Renaux-Petel, Lorenzo Sorbo, Caner \"Unal, and Denis Werth for insightful discussions and comments. 
AC acknowledges funding support from the Initiative Physique des Infinis (IPI), a research training program of the Idex SUPER at Sorbonne Universit\'e. MP acknowledges support from Istituto Nazionale di Fisica Nucleare (INFN) through the Theoretical Astroparticle Physics
(TAsP) project, and from the MIUR Progetti di Ricerca di Rilevante Interesse Nazionale
(PRIN) Bando 2022 - grant 20228RMX4A.

\bibliographystyle{jhep}
\bibliography{main}

\appendix

\section{Analytic computations of the axion perturbations}
\label{app:sigma}

This appendix presents the analytic computation of the power spectrum of the axion field perturbations sourced by the gauge field. The method follows that of Ref.~\cite{Namba:2015gja}, which studied the power spectrum of the inflaton field (sourced by the axion) but not that of the axion.

Decomposing the axion field into a classical homogeneous value plus a quantum operator
\begin{equation}
\sigma \left( \tau ,\, \vec{x} \right) = \sigma \left( t \right) + \delta \sigma \left( \tau ,\, \vec{x} \right) = \sigma \left( t \right) + \int \frac{d^3 k}{\left( 2 \pi \right)^{3/2}} \, {\rm e}^{i \vec{k} \cdot \vec{x}} \, \frac{Q_\sigma \left( \tau ,\, \vec{k} \right)}{a \left( \tau \right)},
\end{equation}
and analogously for the inflaton, the operator $Q_\sigma \left( \tau ,\, \vec{k} \right)$ satisfies the second of eqs.~(\ref{eq:lin}). Disreegarding the backreaction from the inflaton perturbations (namely, neglecing the term proportional to $Q_\phi$ in that equation) and to leading order in slow roll, this equation becomes (cf. eq. (3.10) of~\cite{Namba:2015gja} for details)
\begin{equation}
\left( \frac{\partial^2}{\partial \tau^2} + k^2 - \frac{2}{\tau^2} \right) Q_\sigma \left( \tau ,\, \vec{k} \right) \simeq \frac{\alpha a^3}{f} \int \frac{d^3 x}{\left( 2 \pi \right)^{3/2}} {\rm e}^{-i \vec{k} \cdot \vec{x}} \vec{E} \cdot \vec{B} \equiv S_\sigma \left( \tau ,\, \vec{k} \right).
\label{eq-Qphi}
\end{equation}
where the explicit form of the source can be found in eq. (C.2) of~\cite{Namba:2015gja}. 

The axion field in turn sources inflaton perturbations through the first of eqs.~(\ref{eq:lin}). Ref.~\cite{Namba:2015gja} computed the power spectrum and the bispectrum of the sourced inflaton perturbatons, and fitted their frequency-dependence amplitude as a function of the model parameter $\xi_*$ (for the two choices $\delta = 0.2 ,\, 0.5$). Here we present the results for the computation of the axion power spectrum, which was not given in~\cite{Namba:2015gja}. We are interested in the spectrum at super-horizon scales, which we parametrize as
\begin{equation}
\lim_{k \tau \to 0} \left\langle \delta \sigma \left( \tau ,\, \vec{k} \right) \delta \sigma \left( \tau ,\, \vec{k}' \right) \right\rangle \equiv f^2 \frac{2 \pi^2}{k^3} {\cal P}_{\delta \sigma/f} \left( k \right) \delta^{(3)} \left( \vec{k} + \vec{k}' \right).
\label{Psig}
\end{equation}

Eq.~(\ref{eq-Qphi}) is formally solved by
\begin{equation}
Q_{\sigma,{\rm sources}} \left( \tau ,\, \vec{k} \right) = \int^\tau G_k \left( \tau ,\, \tau' \right) S_\sigma \left( \tau ,\, \vec{k} \right),
\label{Qs-sol}
\end{equation}
where the (de Sitter) Green function in the super-horizon limit is
\begin{equation}
\lim_{k \tau \to 0} G_k \left( \tau ,\, \tau' \right) = \frac{k \tau' \, \cos \left( k \tau' \right) - \sin \left( k \tau' \right)}{k^3 \tau \tau'} \theta \left( \tau - \tau' \right).
\end{equation}

Inserting the solution (\ref{Qs-sol}) into~\cref{Psig}, and evaluating the expectation value as done in appendix C of~\cite{Namba:2015gja} for the power spectrum of the inflaton, we arrive to
\begin{align}
{\cal P}_{\delta \sigma/f} \left( k \right) =& \frac{\alpha^2 H^4}{16 \pi^2 f^4} \int \frac{d^3 {\tilde p}}{\left( 2 \pi \right)^3} \left\vert \epsilon_i^{(+)} \left( \hat{\tilde p} \right) 
\epsilon_i^{(+)} \left( \frac{{\hat k} - \vec{\tilde p}}{\left\vert {\hat k} - \vec{\tilde p} \right\vert} \right) \right\vert^2 {\tilde p}^{1/2} \left\vert {\hat k} - \vec{\tilde p} \right\vert^{1/2} \left( {\tilde p}^{1/2} + \left\vert {\hat k} - \vec{\tilde p} \right\vert^{1/2} \right)^2 \nonumber\\
& N^2 \left[ \xi_* ,\, {\tilde p} x_* ,\, \delta \right] N^2 \left[ \xi_* ,\, \left\vert {\hat k} - \vec{\tilde p} \right\vert x_* ,\, \delta \right] {\cal T}_\sigma^2 \left[ \xi_* ,, x_* ,\, \delta ,\, \sqrt{\tilde p} + \sqrt{\left\vert {\hat k} - \vec{\tilde p} \right\vert} \right] \;,
\end{align}
where we have introduced the rescaled internal momentum $\vec{\tilde p} \equiv \frac{\vec{p}}{k}$ and the rescaled time $x_* \equiv - k \tau_*$ at which the axion roll is fastest. We recall that $N[\xi_*,k,\delta]$ is the normalizaton of the gauge mode functions defined in eq.~(\ref{A-wkb}), while 
\begin{equation}
{\cal T}_\sigma \left[ \xi_* ,\, x_* ,\, \delta ,\, Q \right] \equiv \int_0^\infty d x \left[ \sin x - x \, \cos x \right] {\rm exp} \left[ - \frac{4 \xi_*^{1/2}}{1+\delta} \, \frac{x^{\left(1+\delta\right)/2}}{x_*^{\delta/2}} \, Q \right] \;, 
\end{equation}
arises from the time integration of the Green function times the gauge mode function. Finally, $\vec{\epsilon}^{(+)}$ is the polarization operator of the enhanced gauge mode, and the product entering in this expression is given after eq.~ (C.17) of ~\cite{Namba:2015gja}. 

One angular integration (namely, the rotation of $\vec{p}$ around $\vec{k}$) is trivial; for the remaining two integrations, we introduce the variables $x \equiv {\tilde p} + \left\vert {\hat k} - \vec{\tilde p} \right\vert$ and $y \equiv {\tilde p} - \left\vert {\hat k} - \vec{\tilde p} \right\vert$, and note that the integrand is even in $y$, so that 
\begin{align}
{\cal P}_{\delta \sigma/f} \left( k \right) =& \frac{\alpha^2 H^4}{256 \pi^4 f^4} \int_1^\infty d x \int_{-1}^1 d y \,   \frac{\left(x^2-1\right)^2 \left( \sqrt{x+y} + \sqrt{x-y} \right)^2}{\sqrt{x+y} \, \sqrt{x-y}} \nonumber\\
& N^2 \left[ \xi_* ,\, \frac{x+y}{2} \, x_* ,\, \delta \right] N^2 \left[ \xi_* ,\, \frac{x-y}{2} \, x_* ,\, \delta \right] {\cal T}_\sigma^2 \left[ \xi_* ,, x_* ,\, \delta ,\, \frac{\sqrt{x+y}+\sqrt{x-y}}{\sqrt{2}} \right] \;,
\label{P-ds}
\end{align}

As in~\cite{Namba:2015gja}, we consider the two values $\delta = 0.2,\, 0.5$. We then consider the set of values $\left\{ 3 ,\, 3.5 ,\, 4 ,\, \dots ,\, 7 \right\}$ for $\xi_*$. For each choice of these two parameters we evaluated the rescaled power spectrum
\begin{equation}
\tilde{\cal P}_{\delta \sigma/f} \left( \frac{k}{k_*} ,\, \xi_* ,\, \delta \right) \equiv {\cal P}_{\delta \sigma/f} \left( k \right) \Big/ \frac{\alpha^2 H^4}{f^4} \,.
\end{equation}
where $k_* \equiv a \left( \tau_* \right) H \left( \tau_* \right)$ is the wavenumber of the mode that exited the horizon at the time $\tau_*$, when the roll of the axion is fastest.

\begin{figure}
\centering
\begin{tikzpicture}
\node (img) at (-6.,0){\includegraphics[width=7.cm]{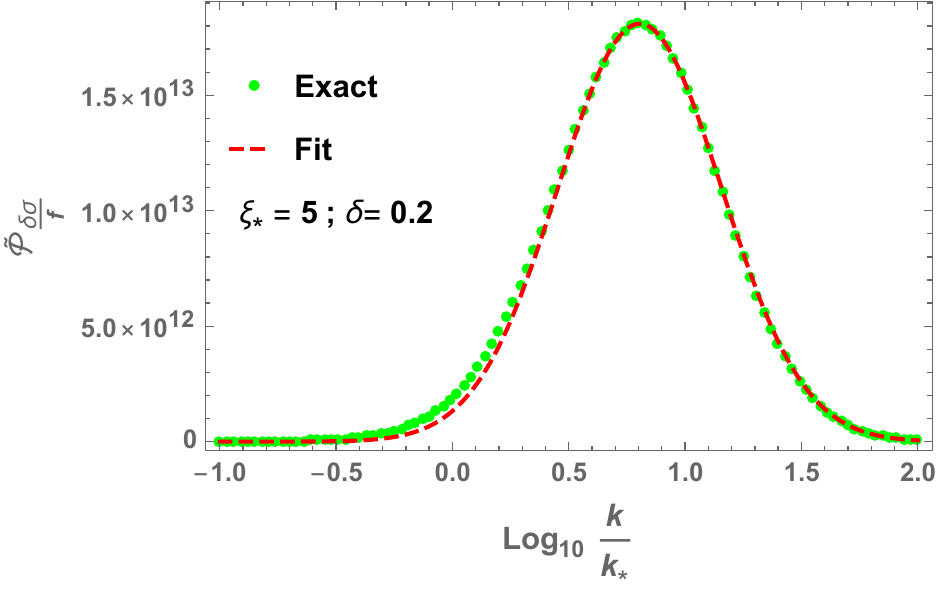}};
\node (img2) at (1.3,0) {\includegraphics[width=7.cm]{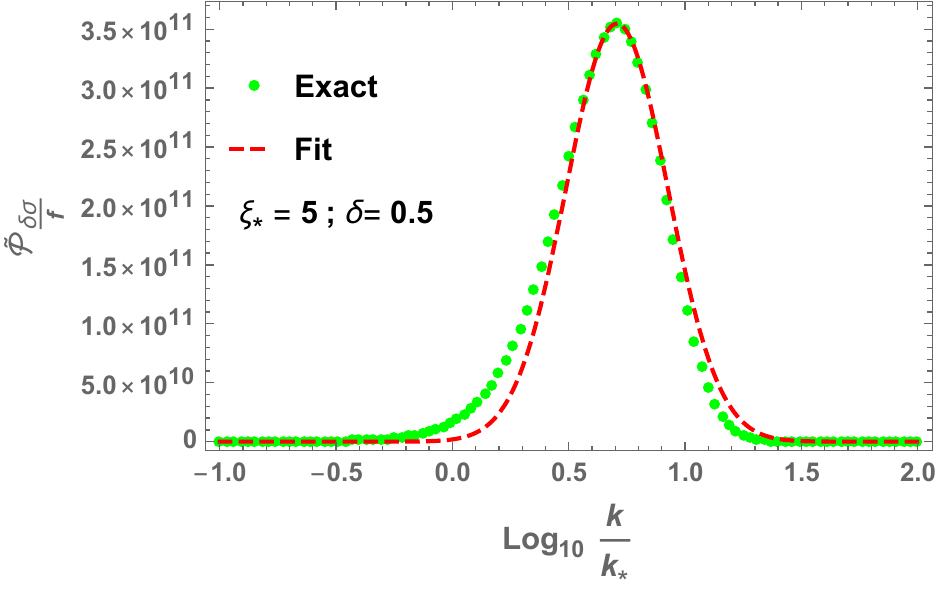}};
\end{tikzpicture}
\caption{Bump of the (rescaled) power spectrum of the source perturbations of the axion field, for $\xi_* = 5$ and for $\delta = 0.2$ (left panel) and $\delta = 0.5$. The exact integration of~(\ref{P-ds}) is compared with the fitting function~(\ref{P-ds-fit}).}
\label{fig:Pds}
\end{figure}

In Figure \ref{fig:Pds} we show the rescaled power
for $\xi_* = 5$ and for the two choices $\delta = 0.2$ (left panel) and $\delta = 0.5$ (right panel). The exact numerical integration of eq.~(\ref{P-ds}), shown with green dots, is compared with the fitting function 
\begin{equation}
\tilde{\cal P}_{\delta \sigma/f} \left( \frac{k}{k_*} ,\, \xi_* ,\, \delta \right) \simeq f_{2,\sigma}^c \left[ \xi_* ,\, \delta \right] \, {\rm exp} \left[ - \frac{1}{2 \sigma_{2,\sigma}^2 \left[ \xi_* ,\, \delta \right]} \, \ln^2 \left( \frac{k}{k_* \, x_{2,\sigma}^c \left[ \xi_* ,\, \delta \right]} \right) \right] \,,
\label{P-ds-fit}
\end{equation}
shown with a red dashed line in the figure. We verifued that the fitting function provides and equally good fit of the exct result for all the other values of $\xi_*$ that we have considered. The function (\ref{P-ds-fit}) is a log-normal bump centered at wavenumbers close to $k_*$, and characterized by three parameters  $f_{2,\sigma}^c ,\, x_{2,\sigma}^c$, and $\sigma_{2,\sigma}^2$, which control, respectively, the height, the central position, and the width of the bump. 

\begin{figure}
\centering
\begin{tikzpicture}
\node (img) at (-5.,0) {\includegraphics[width=5.cm]{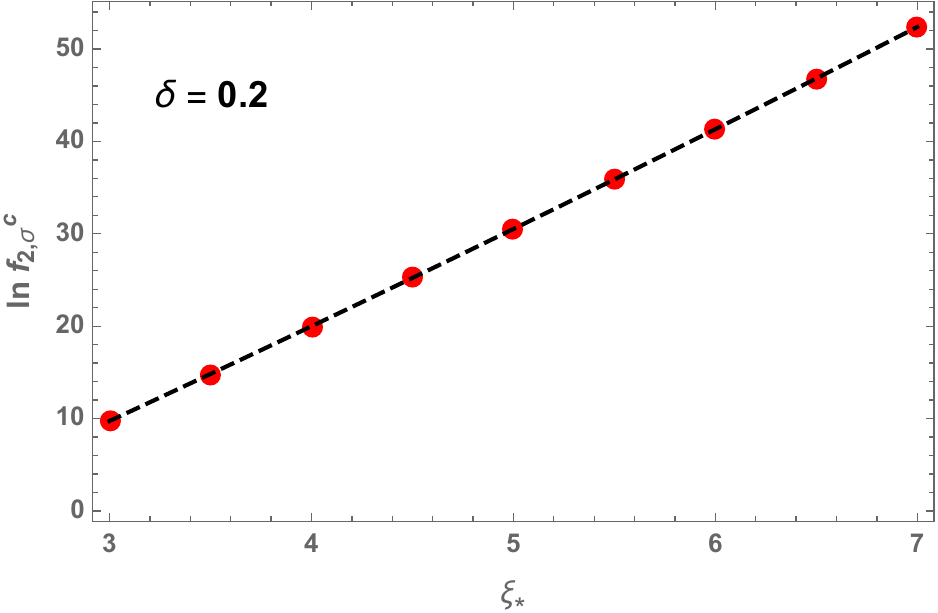}};
\node (img2) at (0,0) {\includegraphics[width=5.cm]{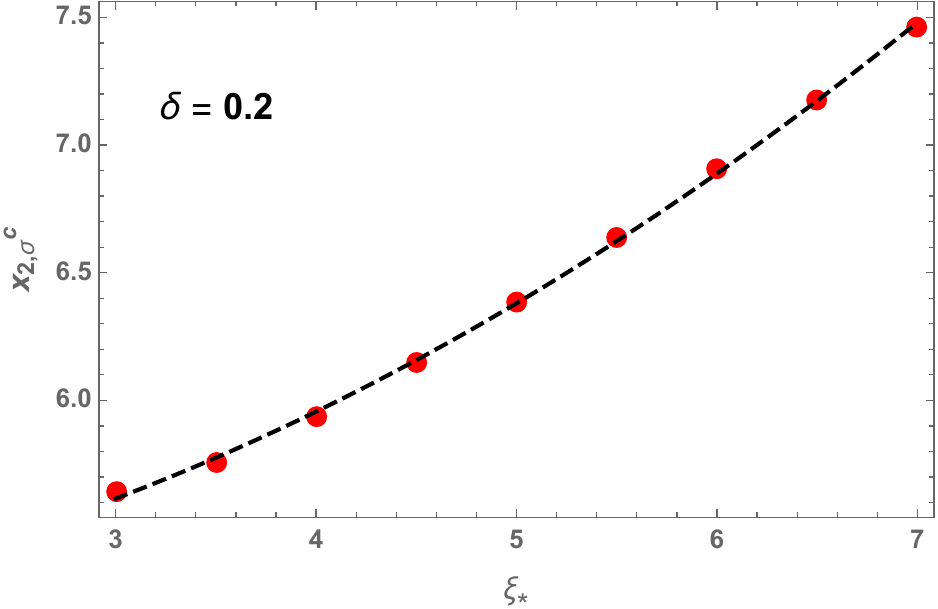}};
\node (img3) at (5.,0) {\includegraphics[width=5.cm]{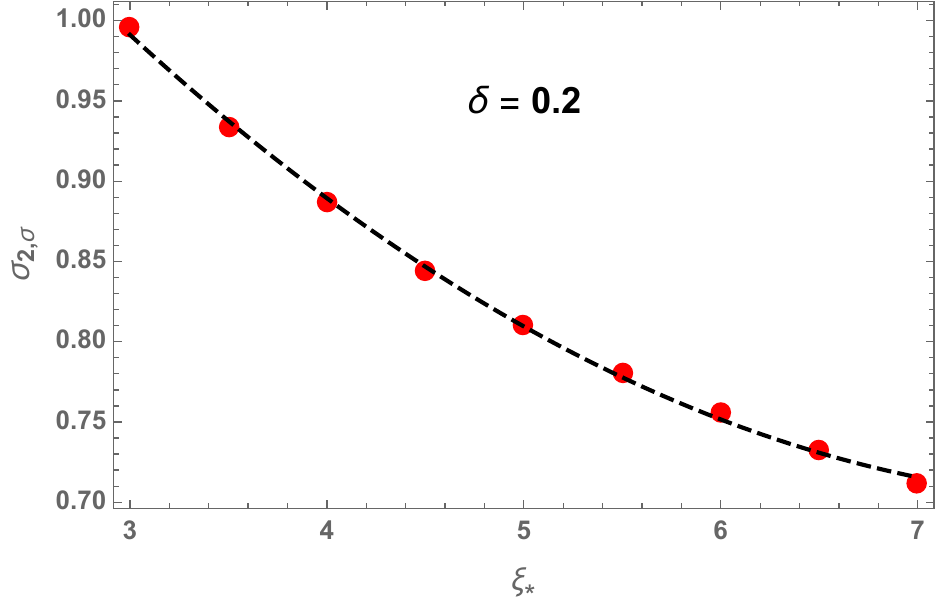}};
\end{tikzpicture}
\caption{Dependence on $\xi_*$ of the fitting parameters $\ln f_{2,\sigma}^c ,\, x_{2,\sigma}^c$, and $\sigma_{2,\sigma}^2$ entering in eq.~(\ref{P-ds-fit}) for the choice $\delta = 0.2$. The red dots denotes the values obtained by evaluating and fitting the power spectrum at fixed $\xi_*$. The
solid lines are the polynomial fits reported in Table~\ref{tab:tabella}.}
\label{fig:Pds-fit-param}
\end{figure}

The three fitting parameters are in turn functions of the model parameters $\xi$ and $\delta$. In Figure~\ref{fig:Pds-fit-param} we show the value of the parameters for $\delta = 0.2$ and for the varoius $\xi_*$ that we have considered. The behaviour of these values with $\xi_*$ is compared with a quadratic polynomial fit. The fit works equally well also for the other choice $\delta = 0.5$ that we have studied. We provide the fitting functions in
Table~\ref{tab:tabella}.

\begin{table}[htbp]
\hspace{-0.5cm}
\begin{tabular}{|c||c|c|c|}
\hline
 & $\ln f_{2,\sigma}^c$ & $x_{2,\sigma}^c$ & $\sigma_{2,\sigma}$ \\
\hline
$\delta = 0.2$ & 
$-19.42 + 9.313 \, \xi_* + 0.1349\, \xi_*^2$ &
$5.097 + 0.047 \, \xi_* + 0.04193\, \xi_*^2$ &
$1.428 - 0.1785 \, \xi_* + 0.01096\, \xi_*^2$ \\
\hline
$\delta = 0.5$ & 
$-18.66 + 8.6457 \, \xi_* + 0.08085\, \xi_*^2$ &
$1.666  +  0.67 \, \xi_* + 0.00332\, \xi_*^2$ &
$0.732 - 0.0672 \, \xi_* + 0.00428\, \xi_*^2$ \\
\hline
\end{tabular}
\caption{Dependence on $\xi_*$ of the parameters in (\ref{P-ds-fit}), for the two values of $\delta$ that we have studied.}
\label{tab:tabella}
\end{table}

\section{Numerical convergence and lattice tests}
\label{app:resolution}

In this appendix, we demonstrate the numerical stability of our results.

\subsection{Resolution dependence tests}
In this section, we show that our results do not depend on the resolution by running simulations with different numbers of points $N_{\rm pts}$ and comoving length $L$, corresponding to different IR and UV resolutions. 

\begin{figure}
	\centering
	
	\begin{tikzpicture}
	\node (img) at (1,0) {\includegraphics[width=11cm]{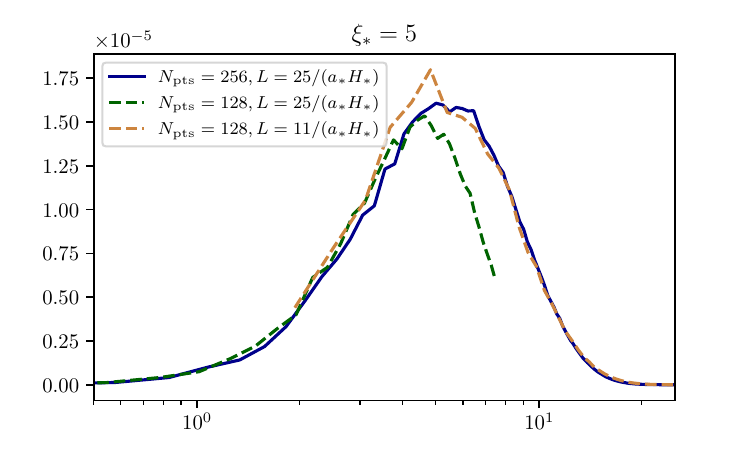}};
	
	\node [rotate=0,text width=1cm,align=center] at (-4.3,0){$\mathcal{P}_{\sigma/f}$ };
   
	\node [text width=0.01cm,align=center] at (.7,-3.3){$k/k_*$};
	\end{tikzpicture}
	
	\caption{Final time power spectrum of $\sigma/f$ in the case of small backreaction ($\xi=5$). Different colors correspond to different resolutions, as given by the legend.}
	\label{fig:psigma_comp}
\end{figure}

\begin{figure}
	\centering
	
	\begin{tikzpicture}
	\node (img) at (1,0) {\includegraphics[width=11cm]{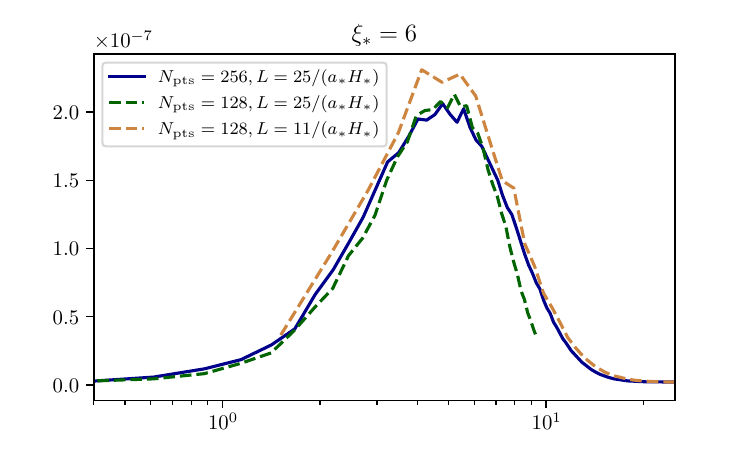}};
	
	\node [rotate=0,text width=1cm,align=center] at (-4.,0){$\mathcal{P}_{\zeta}$ };
    
	\node [text width=0.01cm,align=center] at (1.,-3.3){$k/k_*$};
	\end{tikzpicture}
	
	\caption{Final time power spectrum of $\zeta$ in the case of strong backreaction ($\xi=6 $). Different colors correspond to different resolutions, as given by the legend.}
	\label{fig:pzeta_comp_strong}
\end{figure}

In \cref{fig:psigma_comp} we show the final power spectrum of $\sigma$ in the case of weak backreaction for different lattice resolutions. Similarly, \cref{fig:pzeta_comp_strong} displays the power spectrum of $\zeta$ in the case of strong backreaction. In both cases, we see that the size of the power spectrum does not change with the resolution, indicating that removing IR and UV modes does not affect our results.

\begin{figure}
	\centering
	
	\begin{tikzpicture}
 \node (img2) at (-6.,0) {\includegraphics[scale=.85]{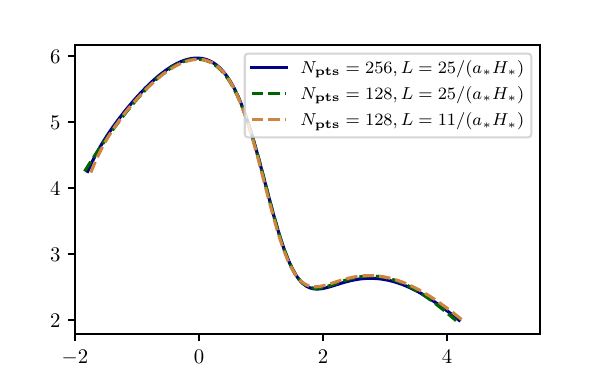}};

 \node [text width=0.01cm,align=center] at (-6,-3.1){$N$};

	\node [rotate=0,text width=1cm,align=center] at (-2.7-7.5,0){$\langle\xi\rangle$};

	\end{tikzpicture}
	
	\caption{Time evolution of $\langle\xi\rangle$ in the case of strong backreaction from simulations with different resolutions, as indicated by the legend.
 }
	\label{fig:hist_pdf}
\end{figure}

In \cref{fig:hist_pdf}, we also plot the evolution of $\xi$ in the case of strong backreaction, showing that the amount of backreaction does not depend on the resolution. 

\subsection{Gauge constraint}
As already mentioned in \cref{sec:eoms}, the Lorenz constraint $ \partial^\mu A_{\mu} = 0 $ is not automatically enforced, as we evolve all four components of the gauge field independently. While an exact solution of the system would respect this constraint, our numerical integration unavoidably introduces small numerical errors that break it. Therefore, we need to check that this breaking of the Lorenz gauge is sufficiently small all throughout the evolution, to ensure that we do not evolve any appreciable spurious effect. 

\begin{figure}
	\centering
	
	\begin{tikzpicture}
	\node (img) at (1,0) {\includegraphics[width=10cm]{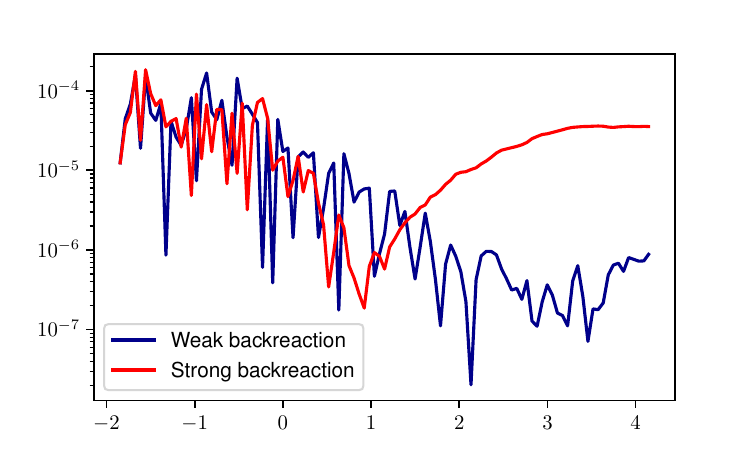}};
	
	\node [rotate=0,text width=1cm,align=center] at (-4.,0){$\mathcal G$ };
    
	\node [text width=0.01cm,align=center] at (1.,-3.2){$N$};
	\end{tikzpicture}
	
	\caption{Time evolution of $\mathcal G$, defined in \cref{eq:gauge_constraint}, quantifying violation of the Lorenz constraint.}
	\label{fig:gauge_constraint}
\end{figure}

To this aim, we introduce the following dimensionless quantity \cite{Deskins_2013,Adshead_2015,Caravano:2022yyv}:
\begin{equation}
\label{eq:gauge_constraint}
\mathcal{G} \equiv \frac{\partial^\mu A_\mu}{\sqrt{\sum_\rho |\partial^\rho A_\rho|^2}},
\end{equation}
which quantifies the violation of the Lorenz constraint in a dimensionless way. In \cref{fig:gauge_constraint}, we show the evolution of this quantity in the cases of strong and weak backreaction. From this plot, we see that the Lorenz constraint is preserved in both cases, and the size of the violation is consistent with what is observed in the minimal model \cite{Caravano:2022epk,Caravano:2022yyv}.

\end{document}